\pdfoutput=1

\documentclass[conference]{IEEEtran}

\usepackage[utf8]{inputenc}
\usepackage[T1]{fontenc}
\usepackage{etex}
\usepackage{microtype}
\usepackage{cite}
\usepackage[cmex10]{amsmath}
\usepackage{amssymb}
\usepackage{mathtools}
\usepackage{array}
\usepackage{booktabs}
\usepackage{tikz}
\usepackage{pgfplotstable}
\usepackage{pgfplots}
\usepackage{xspace}
\usepackage[linesnumbered, algoruled, nokwfunc]{algorithm2e}
\usepackage[inline]{enumitem}

\ifCLASSOPTIONcompsoc
  \usepackage[caption=false,font=normalsize,labelfont=sf,textfont=sf]{subfig}
\else
  \usepackage[caption=false,font=footnotesize]{subfig}
\fi
\usepackage[shadow]{todonotes}
\usetikzlibrary{automata, arrows, calc, decorations.pathreplacing, matrix, shadows}
\tikzfading[name=lefttoright, left color=transparent!0, right color=transparent!75, middle color=transparent!10]
\tikzset{auto, shorten >= 1pt, >=stealth}
\tikzset{initial text={}, initial distance=1em}
\tikzset{every state/.append style={minimum size=3.5ex}}
\tikzset{player0/.style={draw, rectangle, rounded corners=7.5pt}}
\tikzset{player1/.style={draw, rectangle}}
\tikzset{safe/.style={fill=gray!12.5}}

\SetKwFunction{FnInit}{init}
\SetKwFunction{FnOrder}{order}
\SetKwFunction{FnTest}{test}

\SetFuncSty{itshape}

\newtheorem{definition}{Definition}
\newtheorem{theorem}{Theorem}
\newtheorem{lemma}{Lemma}
\newtheorem{example}{Example}
\newtheorem{observation}{Observation}
\newtheorem{corollary}{Corollary}
\renewcommand{\IEEEQED}{\IEEEQEDopen}

\newcommand{\dfa}{DFA\xspace}
\newcommand{\dfas}{DFAs\xspace}
\newcommand{\nfa}{NFA\xspace}
\newcommand{\nfas}{NFAs\xspace}
\newcommand{\myrightarrow}[1]{\xrightarrow{#1}}
\newcommand{\playera}{Player~0\xspace}
\newcommand{\playerb}{Player~1\xspace}
\newcommand{\players}{Player~$\sigma$\xspace}
\newcommand{\vbar}{\ensuremath{\mathsf{l}}}
\newcommand{\Po}{\ensuremath{\mathit{Pos}}\xspace}
\newcommand{\Ne}{\ensuremath{\mathit{Neg}}\xspace}
\newcommand{\Ex}{\ensuremath{\mathit{Ex}}\xspace}
\newcommand{\Un}{\ensuremath{\mathit{Uni}}\xspace}
\newcommand{\true}{\ensuremath{\mathit{true}}\xspace}
\newcommand{\ante}{\ensuremath{\mathit{Ante}}}
\newcommand{\pref}{\ensuremath{\mathit{Pref}}}
\newcommand{\parmmax}{\ensuremath{k'}\xspace}

\newcommand{\defend}{}
\newcommand{\exend}{\hfill\tikz \draw (0, 0) -| +(1ex, 1ex);}

\begin{document}

\author{%
	\IEEEauthorblockN{Daniel Neider}
	\IEEEauthorblockA{%
		Department of Electrical Engineering\\
	 University of California at Los Angeles, USA%
	}
	\and
	\IEEEauthorblockN{Ufuk Topcu}
	\IEEEauthorblockA{%
	Department of Electrical and Systems Engineering\\
		University of Pennsylvania, USA%
	}
}

\title{An Automaton Learning Approach to Solving Safety Games over Infinite Graphs}
\maketitle

\begin{abstract}
We propose a method to construct finite-state reactive controllers for systems whose interactions with their adversarial environment are modeled by infinite-duration two-player games over (possibly) infinite graphs. 
The proposed method targets safety games with infinitely many states or with such a large number of states that it would be impractical---if not impossible---for conventional synthesis techniques that work on the entire state space. We resort to constructing finite-state controllers for such systems through an automata learning approach, utilizing a symbolic representation of the underlying game that is based on finite automata. Throughout the learning process, the learner maintains an approximation of the winning region (represented as a finite automaton) and refines it using different types of counterexamples provided by the teacher until a satisfactory controller can be derived (if one exists). We present a symbolic representation of safety games (inspired by regular model checking), propose implementations of the learner and teacher, and evaluate their performance on examples motivated by robotic motion planning in dynamic environments.
\end{abstract}

\vspace{-0.05in}
\section{Introduction} \vspace{-0.02in}
We propose an automata learning-based method to construct reactive controllers subject to safety specifications. We model the interaction between a controlled system and its possibly adversarial environment as a two-player game over a graph \cite{DBLP:journals/apal/McNaughton93}. We consider games over \emph{infinite graphs}. In this setting, the conventional techniques for reactive controller synthesis (e.g., fixed-point computations) are not applicable anymore. Therefore, we resort to a \emph{learning}-based approach for constructing finite-state reactive controllers for the controlled system. The learning takes place in a setting akin to \emph{counterexample-guided inductive synthesis} (CEGIS)~\cite{DBLP:conf/oopsla/ItzhakyGIS10} between a \emph{teacher}, who has knowledge about the safety game in question, and a \emph{learner}, whose objective is to identify a controller using information disclosed by the teacher in response to (incorrect) conjectures.

A natural context for the proposed method is one in which the interaction between the controlled system and its environment is so complex that it can be represented only by graphs with infinitely many vertices (e.g., motion planning over unbounded grid worlds) or ``practically infinitely many" states (i.e., the number of possible configurations is so large that the game becomes impractical for conventional techniques). Additionally, in situations where a complete description of the game is not available in a format amenable to existing game solvers \cite{SlugsReference, bohy2012acacia+}, there may still exist human experts (or automated oracles, as in Section~\ref{sec:teacher}) who have sufficient insight into how the controlled system should behave and can act as teacher.

We focus on games with safety specifications, which already capture practically interesting properties (e.g., safety and bounded-horizon reachability).
However, games over infinite graphs require special attention on the representation and manipulation of the underlying graph structure. Hence, one of our main contributions is a symbolic representation of safety games, called \emph{rational safety games}, that follows the idea of \emph{regular model checking}~\cite{DBLP:conf/cav/BouajjaniJNT00} in that it represent sets of vertices by regular languages and edges by so-called rational relations.

A straightforward approach to solve (rational) safety games is computing a \emph{winning set} for the controlled system (i.e., a safe subset of the vertices in which the system can force to remain). Once a winning set is computed, a strategy for the system is determined by choosing its moves (in each of its turns) to stay inside the set, which is possible regardless of the moves of the environment. We use winning sets as a proxy for an actual controller, and the objective of the learning task is the construction of a winning set. In fact, learning a winning set rather than a controller results in more permissive strategies (and potentially smaller solutions) as the moves of the system do not need to be fixed during the learning process.

We develop a framework for learning winning sets for rational safety games and particular implementations of a teacher and learner. The actual learning works iteratively. In each iteration, the learner conjectures a winning set, represented as a deterministic finite automaton. The teacher performs a number of checks and returns, based on whether the conjecture passes the checks, a counterexample. Following the ICE learning framework \cite{DBLP:conf/cav/0001LMN14} and partially deviating from the classical learning frameworks for regular languages \cite{DBLP:journals/iandc/Gold78,DBLP:journals/iandc/Angluin87}, the counterexample may be one of the following four types: positive, negative, existential implication and universal implication counterexamples. Based on the response from the teacher, the learner updates his conjecture. If the conjecture passes all checks (i.e., the teacher returns no counterexample), the learning process terminates with the desired controller. 


A learning-based approach offers several advantages: First, even though the underlying game may be prohibitively large, the reactive controller necessary to realize the specifications often has a compact representation in practice; for example, depending on the given task specification in a robotic motion planning scenario, only a small subset of all possible rich interactions between the robot and its dynamic environment over a possibly large workspace is often relevant.
Second, since learning-based approaches usually identify ``small" solutions (as they typically produce intermediate conjectures of increasing size), their runtime mainly depends on the size of the solution rather than the size of the underlying game.
Third, learning-based approaches reduce the gap between human designers and construction of reactive controllers by hiding  the complexity of the underlying game from the learner.

Finally, we demonstrate the use of our overall learning-based framework empirically on a series of examples motivated by robotic motion planning in dynamic environments.\par\vspace{.5\baselineskip}

\subsubsection*{Related Work}
Games over infinite graphs have been studied in the past, predominantly in the case of games over pushdown graphs~\cite{DBLP:conf/birthday/KupfermanPV10}. The games we consider here, however, are played over a richer class of graphs and require different techniques to be solved. Also, a constraint-based approach to solving games over infinite graphs has recently been proposed~\cite{DBLP:conf/popl/BeyeneCPR14}.
 
Learning-based techniques for games over infinite graphs have already been studied in the context of reachability games~\cite{DBLP:conf/wia/Neider10}; in fact, our symbolic representation of safety games is a generalization of the representation proposed there. In the context of safety games, recent work \cite{DBLP:conf/atva/Neider11} has already demonstrated the ability of learning-based approaches to extract small reactive controllers from a~priori constructed controllers with possibly large number of states. In this work, we by-pass the a~priori construction of possibly large reactive controllers by learning (an appropriate representation of) a controller directly.

\section{Rational Safety Games}

This section recaps infinite-duration, two-player safety games as well as basic concepts of automata theory and introduces rational safety games.  

\paragraph{Safety Games}
We consider safety games (i.e., infinite duration two-person games on graphs) as popularized by McNaughton~\cite{DBLP:journals/apal/McNaughton93}. A safety game is played on an \emph{arena} $\mathfrak A = (V_0, V_1, E)$ consisting of two nonempty, disjoint sets $V_0, V_1$ of \emph{vertices} (we denote their union by $V$) and a directed edge relation $E \subseteq V \times V$. In contrast to the classical (finite) setting, we allow $V_0$ and $V_1$ to be countable sets. 
As shorthand notation, we write the successors of a set $X \subseteq V$ of vertices as $E(X) = \{ y \mid \exists x \in X \colon (x, y) \in E \}$.

We consider safety games with initial vertices, which are defined as triples $\mathfrak G = (\mathfrak A, F, I)$ consisting of an arena $\mathfrak A = (V_0, V_1, E)$, a set $F \subseteq V$ of \emph{safe vertices}, and a set $I \subseteq F$ of \emph{initial vertices}. Such safety games are played by two players, named \playera and \playerb, as follows: A token is placed on some initial vertex $v_0 \in I$ and, in each turn, the player owning the current vertex moves the token to a successor vertex of his choice. This process of moving the token is repeated ad infinitum, thereby forming an infinite sequence of vertices, which is called a play. Formally, a \emph{play} is an infinite sequence  $\pi = v_0 v_1 \ldots \in V^\omega$ that satisfies $v_0 \in I$ and $(v_i, v_{i+1}) \in E$ for all $i \in \mathbb N$. The set $F$ defines the winning condition of the game in the sense that a play $v_0 v_1 \ldots$ is \emph{winning for \playera} if $v_i \in F$ for all $i \in \mathbb N$---otherwise it is \emph{winning for \playerb}.

A strategy for \players, $\sigma \in \{ 0, 1 \}$, is a mapping $f_\sigma \colon V^* V_\sigma \to V$, which prescribes how to continue playing. A strategy $f_\sigma$ is called \emph{winning} if any play $v_0 v_1 \ldots$ that is played according to the strategy (i.e., that satisfies $v_{i+1} = f_\sigma(v_0 \ldots v_i)$ for all $i \in \mathbb N$ and $v_i \in V_\sigma$) is winning for Player~$\sigma$. A winning strategy for \playera straightforwardly translates into a controller satisfying the given safety specifications and, hence, we restrict ourselves to compute winning strategies for \playera.

Computing a winning strategy for \playera is usually reduced to finding a so-called winning set.

\begin{definition}[Winning set]\label{main:def:winning_set}
\itshape
For a safety game $\mathfrak G = (\mathfrak A, I, F)$ over the arena $\mathfrak A = (V_0, V_1, E)$, a \emph{winning set} is a set $W \subseteq V$ satisfying
\begin{enumerate*}[label=(\arabic*)]
	\item \label{main:cond:winning_set:1} $I \subseteq W$,
	\item \label{main:cond:winning_set:2} $W \subseteq F$,
	\item \label{main:cond:winning_set:3} $E(\{ v \}) \cap W \neq \emptyset$ for all $v \in W \cap V_0$ (\emph{existential closedness}), and
	\item \label{main:cond:winning_set:4} $E(\{ v \}) \subseteq W$ for all $v \in W \cap V_1$ (\emph{universal closedness}). \defend
\end{enumerate*}
\end{definition}

By computing a winning set, one immediately obtains a strategy for \playera: starting in an initial vertex, \playera simply moves to a successor vertex inside $W$ whenever it is his turn. A straightforward induction over the length of plays proves that every play that is played according to this strategy stays inside $F$, no matter how \playerb plays, and, hence, is won by \playera (since $I \subseteq W \subseteq F$). A winning set is what we want to compute---or, more precisely, \textit{learn}. 

Games over infinite arenas require a symbolic representation in order to work with them algorithmically. We follow the idea of \emph{regular model checking}~\cite{DBLP:conf/cav/BouajjaniJNT00}, an approach in verification, and represent sets of vertices by regular languages and edges by so-called rational relations. Before we can introduce our symbolic representation of safety games, however, we need to recap basic concepts and notations of automata theory.

\paragraph{Basics of Automata Theory} An \emph{alphabet} $\Sigma$ is a nonempty, finite set, whose elements are called \emph{symbols}. A \emph{word} over the alphabet $\Sigma$ is a sequence $u = a_1 \ldots a_n$ of symbols $a_i \in \Sigma$ for $i \in \{1, \ldots, n\}$; the empty sequence is called \emph{empty word} and denoted by $\varepsilon$.
Given two words $u = a_1 \ldots a_m$ and $v = b_1 \ldots b_n$, the \emph{concatenation of $u$ and $v$} is the word $u \cdot v = uv = a_1 \ldots a_m b_1 \ldots b_n$.
The set of all words over the alphabet $\Sigma$ is denoted by $\Sigma^\ast$, and a subset $L \subseteq \Sigma^\ast$ is called a \emph{language}. The set of prefixes of a language $L \subseteq \Sigma^\ast$ is the set $\pref(L) = \{ u \in \Sigma^\ast \mid \exists v \in \Sigma^\ast \colon uv \in L \}$.

A \emph{nondeterministic finite automaton (NFA)} is a tuple $\mathcal A = (Q, \Sigma, q_0, \Delta, F)$ consisting of a nonempty, finite set $Q$ of \emph{states}, an \emph{input alphabet} $\Sigma$, an \emph{initial state} $q_0 \in Q$, a \emph{transition relation} $\Delta \subseteq Q \times \Sigma \times Q$, and a set $F \subseteq Q$ of \emph{final states}. A \emph{run} of an \nfa $\mathcal A$ on a word $u = a_1 \ldots a_n$ is a sequence of states $q_0, \ldots, q_n$ such that $(q_{i-1}, a_i, q_i) \in \Delta$ for $i \in \{ 1, \ldots, n \}$. We denote this run by $\mathcal A \colon q_0 \myrightarrow{u} q_n$. An \nfa $\mathcal A$ \emph{accepts} a word $u \in \Sigma^\ast$ if $\mathcal A \colon q_0 \myrightarrow{u} q$ with $q \in F$.
The set $L(\mathcal A) = \{u \in \Sigma^\ast \mid \mathcal A \colon q_0 \myrightarrow{u} q, q \in F \}$ is called \emph{language of $\mathcal A$}. A language $L$ is said to be \emph{regular} if there exists an \nfa $\mathcal A$ with $L(\mathcal A) = L$. Finally, $\mathit{NFA}_\Sigma$ denotes the set of all \nfas over $\Sigma$.

A \emph{deterministic finite automaton (DFA)} is an NFA in which $(p, a, q) \in \Delta$, $(p, a, q') \in \Delta$ implies $q = q'$. We replace the transition relation $\Delta$ with a transition function $\delta \colon Q \times \Sigma \to Q$.

We define rational relations by resorting to transducers.
A \emph{transducer} is an \nfa $\mathcal T = (Q, \hat{\Sigma}, q_0, \Delta, F)$ over the alphabet $\hat{\Sigma} = (\Sigma \cup \{ \varepsilon \}) \times (\Gamma \cup \{ \varepsilon \})$---$\Sigma$ and $\Gamma$ are both alphabets---that processes pairs $ (u, v) \in \Sigma^\ast \times \Gamma^\ast$ of words. The \emph{run} of a transducer $\mathcal T$ on a pair $(u, v)$ is a sequence $q_0, \ldots, q_n$ of states such that $(q_{i-1}, (a_i, b_i), q_i) \in \Delta$ for all $i \in \{1, \ldots, n \}$, $u = a_1 \ldots a_n$, and $v = b_1 \ldots b_n$; note that $u$ and $v$ do not need to be of equal length since any $a_i$ or $b_i$ can be $\varepsilon$. A pair $(u, v)$ is said to be \emph{accepted} by $\mathcal T$ if there exists a run of $\mathcal T$ on $(u, v)$ that starts in the initial state and ends in a final state.
As an acceptor of pairs of words, a transducer $\mathcal T$ \emph{defines} a relation, namely the relation consisting of exactly the pairs accepted by $\mathcal T$, which we denote by $R(\mathcal T)$.
Finally, a relation $R \subseteq \Sigma^\ast \times \Gamma^\ast$ is called \emph{rational} if there exists a transducer $\mathcal T$ with $R(\mathcal T) = R$.
(This definition of rational relations is simplified from that in~\cite{DBLP:journals/mst/BlumensathG04} but sufficient for our purpose.)

Our learning framework relies on the two well-known facts. 

\begin{lemma}\label{main:lem:transducer}
\itshape
Let $R \subseteq \Sigma^\ast \times \Gamma^\ast$ be a rational relation and $X \subseteq \Sigma^\ast$ a regular set. Then,
\begin{enumerate*}[label=(\arabic*), ref=\arabic*]
	\item \label{main:itm:lem:transducer:1} the relation $R^{-1} = \{ (y, x) \mid (x, y) \in R \}$ is again rational, and a transducer defining this set can be constructed in linear time; and
	\item \label{main:itm:lem:transducer:2} the set $R(X) = \{ y \in \Gamma^\ast \mid \exists x \in X \colon (x, y) \in R \}$, called the \emph{image of $X$ under $R$}, is again regular, and an \nfa accepting this set can be constructed effectively. 
\end{enumerate*}
\end{lemma}

\paragraph{Rational Safety Games}
A rational safety game is a symbolic representation of a safety game in terms of  regular languages and rational relations.

\begin{definition}
\itshape
A \emph{rational arena} over the alphabet $\Sigma$ is an arena $\mathfrak A_\Sigma = (V_0, V_1, E)$ where $V_0, V_1 \subseteq \Sigma^\ast$ are regular languages  and $E \subseteq V \times V$ is a rational relation. \defend
\end{definition}

The definition of rational safety games is now immediate.

\begin{definition}
\itshape
A \emph{rational safety game} over the alphabet $\Sigma$ is a safety game $\mathfrak G_\Sigma = (\mathfrak A_\Sigma, F, I)$ where $\mathfrak A_\Sigma$ is a rational arena over $\Sigma$ and  $F, I \subseteq \Sigma^\ast$ are regular languages. \defend
\end{definition}

In the remainder, we assume regular languages to be given as \nfas and rational relations as transducers. In addition, we use these notions interchangeably when referring to rational arenas and rational safety games; for instance, we write a rational area $\mathfrak A_\Sigma = (V_0, V_1, E)$ as $\mathfrak A_\Sigma = (\mathcal A_{V_0}, \mathcal A_{V_1}, \mathcal T_E)$ given that $L(\mathcal A_{V_0}) = V_0$, $L(\mathcal A_{V_1}) = V_1$, and $R(\mathcal T_E) = E$.

Let us illustrate rational safety games through an example.

\begin{example}\label{main:ex:rational_safety_game}
Consider a simple example motivated by motion planning, sketched in Figure~\ref{main:fig:example:robot}, in which a robot moves on an infinite, discrete one-dimensional grid that is ``bounded on the left''.
The robot can move left or right to an adjacent cell (provided that it has not reached edge of the grid) or it can stay at its current position. The grid is partitioned into a safe and an unsafe area, the former being shown shaded in Figure~\ref{main:fig:example:robot}. The safe area is parameterized by an integer $k \in \mathbb N \setminus \{ 0 \}$ and consists of all position greater than or equal to $k$. The robot starts somewhere inside the safe area.


\newsavebox{\robot}
\savebox{\robot}{
		\begin{tikzpicture}[scale=.9]
			\draw[thick, fill=white] (0, 0, 0) -- ++(0, -1.2, 0) [draw] arc (180:360:.6 and .15) -- +(0, 1.2, 0) arc (0:360:.6 and .15);
			\draw[thick, fill=white]  (.6, .5, 0) circle (.45);
			\shade[ball color=blue!10!white,opacity=.3] (.6, .5, 0) circle (.45);
			\fill[black] (-.15, -.25, 0) rectangle +(.05, -.8, 0);
			\fill[black] (1.35, -.25, 0) rectangle +(-.05, -.8, 0);
			\fill[black] (.08, -1.5, 0) rectangle +(.4, -.1, 0);
			\fill[black] (1.12, -1.5, 0) rectangle +(-.4, -.1, 0);
		\end{tikzpicture}
}

\begin{figure}
	\centering
	\subfloat[A robot moving on a one-dimensional discrete grid. The figure shows the setting for $k = 2$. \label{main:fig:example:robot}] {
	\begin{tikzpicture}

		\fill[white, drop shadow] (0, 0, 0) -- ++(7, 0, 0) -- ++(0, 0, 1) -- ++(-7, 0, 0) -- cycle;

		\fill[safe] (2, 0, 0) -- ++(4, 0, 0) -- ++(0, 0, 1) -- ++(-4, 0, 0) -- cycle;
		\fill[safe, path fading=lefttoright] (6, 0, 0) -- ++(1, 0, 0) -- ++(0, 0, 1) -- ++(-1, 0, 0) -- cycle;

		\foreach \i in {1, 2, ..., 6} {
			\draw (\i, 0, 0) -- ++(0, 0, 1);
		}
		\draw (6.25, 0, 0) -- ++(-6.25, 0, 0) -- ++(0, 0, 1) -- ++(6.25, 0, 0);
		\draw[dashed] (6.25, 0, 0) -- +(.75, 0, 0);
		\draw[dashed] (6.25, 0, 1) -- +(.75, 0, 0);

		\draw[fill=white, drop shadow={shadow xshift=.75pt, shadow yshift=-.75pt,}] (3.95, .02, .5) -- ++(0, 0, .15) -- ++(.35, 0, 0) -- ++(0, 0, .2) -- ++(.2, 0, -.35) -- ++(-.2, 0, -.35) -- ++(0, 0, .2) -- ++(-.35, 0, 0) -- cycle;
		\draw[fill=white, rotate around={180:(3.05, .02, .5)}, drop shadow={shadow xshift=.75pt, shadow yshift=-.75pt,}] (3.05, .02, .5) -- ++(0, 0, .15) -- ++(.35, 0, 0) -- ++(0, 0, .2) -- ++(.2, 0, -.35) -- ++(-.2, 0, -.35) -- ++(0, 0, .2) -- ++(-.35, 0, 0) -- cycle;

		\node[anchor=south, scale=.5, inner sep=0pt] at (3.5, 0, 0.5) {\usebox{\robot}};

		\node[xslant=.85] at (.5, 0, -.5) {$0$};
		\node[xslant=.85] at (1.5, 0, -.5) {$1$};
		\node[xslant=.85] at (2.5, 0, -.5) {$2$};
		\node[xslant=.85] at (4.5, 0, -.5) {$4$};
		\node[xslant=.85] at (5.5, 0, -.5) {$5$};
		\node[xslant=.85] at (6.5, 0, -.5) {$\ldots$};

	\end{tikzpicture}
	}

	\subfloat[The safety game $\mathfrak G_2$. \playera vertices are drawn as ellipses and and \playerb vertices are drawn as squares. Shaded vertices belong to $F$.\label{main:fig:example:safety_game}] {
	\begin{tikzpicture}[font=\small]	
		\begin{scope}[minimum width=3.25em]
			\node[player0] (s-0) at (0, 0) {$(s, 0)$};
			\node[player1, right=2.25em] (e-1) at (s-0.east) {$(e, 1)$};
			\node[player0, safe, right=2.25em] (s-2) at (e-1.east) {$(s, 2)$};
			\node[player1, safe, right=2.25em] (e-3) at (s-2.east) {$(e, 3)$};
			\node[player1, draw=none, overlay, right=2.25em] (s-4) at (e-3.east) {};

			\node[player1, below=2em] (e-0) at (s-0.south) {$(e, 0)$};
			\node[player0, right=2.25em] (s-1) at (e-0.east) {$(s, 1)$};
			\node[player1, safe, right=2.25em] (e-2) at (s-1.east) {$(e, 2)$};
			\node[player0, safe, right=2.25em] (s-3) at (e-2.east) {$(s, 3)$};
			\node[player1, draw=none, overlay, right=2.25em] (e-4) at (s-3.east) {};
		\end{scope}	
		\node[inner sep=0pt] at ([xshift=-.25em]e-4) {\dots};

		\foreach \pos  in {0, 1, ..., 3} {
			\pgfmathtruncatemacro{\nextpos}{\pos + 1}
			\draw[->] (s-\pos) edge[bend left=15]  (e-\nextpos);
			\draw[->] (e-\nextpos) edge[bend left=15]  (s-\pos);
		}
		\foreach \col  in {0, 1, ..., 3} {
			\draw[->] (e-\col) edge[bend left=15]  (s-\col);
			\draw[->] (s-\col) edge[bend left=15]  (e-\col);
		}
	\end{tikzpicture}
	}
	\caption{Illustration of the safety game discussed in the introductory example.} 
\end{figure}
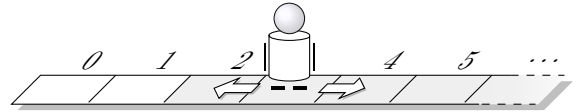
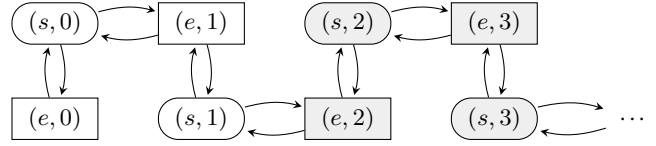 

The robot's movement is governed by two adversarial players, called \emph{system} and \emph{environment}; the system can move the robot to the right or keep it at its current position, whereas the environment can move the robot to the left (if the edge has not been reached) or keep it at its current position. The players move the robot in alternation, and the system moves first. The system's objective is to stay within the safe area, whereas the environment wants to move the robot out of it. Note that the system can win, irrespective of $k$, by always moving right.

A formalization as safety game is straightforward. \playera corresponds to the system and \playerb corresponds to the environment. The arena $\mathfrak A = (V_0, V_1, E)$ consists of vertices $V_0 = \{ s \} \times \mathbb N$ and $V_1 = \{ e \} \times \mathbb N$---$s$, respectively $e$, indicates the player moving next---as well as the edge relation $E = \bigl\{ \bigl((s, i), (e, i+1) \bigr) \mid i \in \mathbb N \bigr\} \cup \bigl\{ \bigl((e, i+1), (s, i) \bigr) \mid i \in \mathbb N \bigr\}$. The safety game itself is the triple $\mathfrak G_k = (\mathfrak A, F, I)$ with $F = \{ s, e \} \times \{ i \in \mathbb N \mid i \geq k \}$ and $I = \{ s \} \times \{ i \in \mathbb N \mid i \geq k \}$. Figure~\ref{main:fig:example:safety_game} sketches the game $\mathfrak G_k$ for the case $k=2$.

We now turn $\mathfrak G_k$ into a rational safety game. To this end, we label each vertex uniquely with a finite word. In our example, we choose $\Sigma =\{s, e, \vbar \}$ and associate the vertex $(x, i) \in \{ s, e \} \times \mathbb N$ with the word $x\vbar^i$ where $\vbar^i$ is the encoding of $i$ in unary. We represent the sets $V_0$ and $V_1$ by the following \nfas: 

\vskip 0.075in
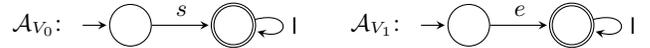
\begin{center}
	\begin{tikzpicture}[node distance=2.25em, font=\small]
		\node[state, initial] (q0) at (0, 0) {};
		\node[state, right=of q0, accepting] (q1) {};
		\node[anchor=east] at ([xshift=-2.25em] q0) {$\mathcal A_{V_0}$:};
		\draw[->] (q0) edge node {$s$} (q1);
		\draw[->] (q1) edge[loop right] node {$\vbar$} ();
	\end{tikzpicture}
	\hskip 1em
	\begin{tikzpicture}[node distance=2.25em, font=\small]
		\node[state, initial] (q0) at (0, 0) {};
		\node[state, right=of q0, accepting] (q1) {};
		\node[anchor=east] at ([xshift=-2.25em] q0) {$\mathcal A_{V_1}$:};
		\draw[->] (q0) edge node {$e$} (q1);
		\draw[->] (q1) edge[loop right] node {$\vbar$} ();
	\end{tikzpicture}
\end{center}
Moreover, we represent the edges by the following transducer: 
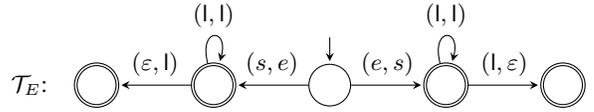
\begin{center}
	\begin{tikzpicture}[node distance=2.75em, font=\small]
		\node[state] (q0) at (0, 0) {};
		\node[state, accepting, left= of q0] (q1) {};
		\node[state, accepting, left= of q1] (q2) {};
		\node[state, accepting, right= of q0] (q3) {};
		\node[state, accepting, right= of q3] (q4) {};
		\node[anchor=east] at ([xshift=-1.5em] q2) {$\mathcal T_E$:};
		\draw[<-, shorten > = 0pt, shorten < = 1pt] (q0.north) -- +(0, 1em);
		\draw[->] (q0) edge node[swap] {$(s, e)$} (q1) edge node {$(e, s)$} (q3);
		\draw[->] (q1) edge[loop above] node {$(\vbar, \vbar)$} () edge node[swap] {$(\varepsilon, \vbar)$} (q2);
		\draw[->] (q3) edge[loop above] node {$(\vbar, \vbar)$} () edge node {$(\vbar, \varepsilon)$} (q4);
\end{tikzpicture}
\end{center}
Finally, the \nfa
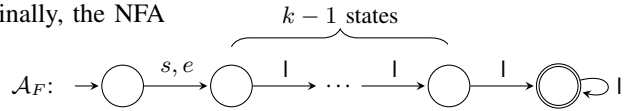
\begin{center}
	\begin{tikzpicture}[node distance=2.5em, font=\small]
		\node[state, initial] (q0) at (0, 0) {};
		\node[state, right=of q0] (q1) {};
		\node[state, draw=none, right=of q1] (q2) {};
		\node[state, right=of q2] (q3) {};
		\node[state, accepting, right=of q3] (q4) {};
		\node at (q2) {\dots};
		\node[anchor=east] at ([xshift=-2.25em] q0) {$\mathcal A_F$:};
		\draw[->] (q0) edge node {$s, e$} (q1);
		\draw[->] (q1) edge node {$\vbar$} (q2);
		\draw[->] (q2) edge node {$\vbar$} (q3);
		\draw[->] (q3) edge node {$\vbar$} (q4);
		\draw[->] (q4) edge[loop right] node {$\vbar$} ();
		\draw[decorate, decoration={brace, amplitude=1.5ex}] ([yshift=1.5ex] q1.north) -- node[overlay, yshift=1ex] {$k-1$ states} ([yshift=1.5ex] q3.north);
	\end{tikzpicture}
\end{center}
represents the set $F$; similarly, $I$ is represented by a copy of $\mathcal A_F$ in which the transition labeled with $e$ is omitted. \exend
\end{example}

It is worth mentioning that rational arenas not only subsume finite arenas but also a rich class of of infinite arenas, including such encoding computations of Turing machines. Hence, the problem of determining the winner of a rational safety game is undecidable, and any algorithm for computing a winning set can at best be a semi-algorithm (i.e., an algorithm that, on termination, gives the correct answer but does not guarantee to halt). The algorithm we design in this paper is of this kind and guarantees to learn a winning set if one exists. To ease description, we always assume that a winning set set exists.


\section{The Learning Framework}
Our learning framework is an extension of the ICE framework proposed by Garg et.~al.~\cite{DBLP:conf/cav/0001LMN14}, which deals with learning loop invariants from positive and negative data as well as implications.
The learning takes place between a \emph{teacher}, who has (explicit or implicit) knowledge about the rational safety game in question, and a \emph{learner}, whose objective is to learn a \dfa accepting a winning set, but who is agnostic to the game. We assume that the teacher announces the alphabet of the game before the actual learning starts.

The learning proceeds in a CEGIS-style loop~\cite{DBLP:conf/oopsla/ItzhakyGIS10}. In every iteration, the learner conjectures a \dfa, let us call it $\mathcal C$, and the teacher checks whether $L(\mathcal C)$ is a winning set---this kind of query is often called \emph{equivalence} or \emph{correctness query}. Although the teacher does not know a winning set (the overall objective is to learn one after all), he can resort to Conditions~\ref{main:cond:winning_set:1}--\ref{main:cond:winning_set:4} of Definition~\ref{main:def:winning_set} in order to decide whether $L(\mathcal C)$ is a winning set. 
If $L(\mathcal C)$ satisfies Conditions~\ref{main:cond:winning_set:1}--\ref{main:cond:winning_set:4} (i.e., $L(\mathcal C)$ is a winning set), then the teacher replies ``yes’’ and the learning ends. If this is not the case, the teacher returns a \emph{counterexample} witnessing the violation of one of these conditions, and the learning continues with the next iteration. The definition below fixes the protocol between the teacher and the learner and defines counterexamples.


\begin{definition}[Teacher for rational safety games] \label{main:def:teacher}
\itshape
Let $\mathfrak G_\Sigma = (\mathfrak A_\Sigma, F, I)$ be a rational safety game over the rational arena $\mathfrak A_\Sigma = (V_0, V_1, E)$. Confronted with a \dfa $\mathcal C$, a \emph{teacher for $\mathfrak G_\Sigma$} replies as follows:
\begin{enumerate}
	\item \label{main:check:teacher:1} If $I \not\subseteq L(\mathcal C)$, then the teacher returns a \emph{positive counterexample} $u \in I \setminus L(\mathcal C)$.
	\item \label{main:check:teacher:2} If $L(\mathcal C) \not\subseteq F$, then the teacher returns a \emph{negative counterexample} $u \in L(\mathcal C) \setminus F$.
	\item \label{main:check:teacher:3} If there exists $u \in L(\mathcal C) \cap V_0$ such that $E(\{u\}) \cap L(\mathcal C) = \emptyset$, then the teacher picks such a $u$ and returns an \emph{existential implication counterexample} $(u, \mathcal A) \in \Sigma^\ast \times \mathit{NFA}_\Sigma$ where $L(\mathcal A) = E(\{u\})$.
	\item \label{main:check:teacher:4} If there exists $u \in L(\mathcal C) \cap V_1$ such that $E(\{u\}) \not\subseteq L(\mathcal C)$, then the teacher picks such a $u$ and returns a \emph{universal implication counterexample} $(u, \mathcal A) \in \Sigma^\ast \times \mathit{NFA}_\Sigma$ where $L(\mathcal A) = E(\{u\})$.
\end{enumerate}
If $\mathcal C$ passes all four checks, the teacher replies ``yes’’. The order in which the teacher performs these checks is arbitrary. \defend
\end{definition}

It is easy to see that the language of a conjecture is indeed a winning set if the teacher replies ``yes’’ (since it satisfies all conditions of Definition~\ref{main:def:winning_set}).
The meaning of a positive counterexample is that any conjecture needs to accepts it, but it was rejected. Similarly, a negative counterexample indicates that any conjecture has to reject it but it was  accepted. An existential implication counterexample $(u, \mathcal A)$ means that any conjecture accepting $u$ has to accept at least one $v \in L(\mathcal A)$, which was violated by the current conjecture. Finally, a universal implication counterexample $(u, \mathcal A)$ means that any conjecture accepting $u$ needs to accept all $v \in L(\mathcal A)$. At this point, it is important to note that Definition~\ref{main:def:teacher} is sound (in particular, both types of implication counterexamples are well-defined due to Lemma~\ref{main:lem:transducer} Part \ref{main:itm:lem:transducer:2}) and every counterexample is a finite object.

Let us illustrate this learning framework through an example.

\begin{example}
We revisit the setting of Example~\ref{main:ex:rational_safety_game} for the case $k = 2$ and describe how the learner learns a winning set. 

Suppose that the learner conjectures the \dfa $\mathcal C_0$ with $L(\mathcal C_0) = \emptyset$. As $\mathcal C_0$  fails Check~\ref{main:check:teacher:1} (it passes all other checks), the teacher returns a positive counterexample, say $u = s\vbar\vbar \in I$.

Next, suppose the learner conjectures the \dfa $\mathcal C_1$ with $L(\mathcal C_1) = \{ s \vbar^n \mid n \geq 2\}$, which passes all checks but Check~\ref{main:check:teacher:3} (as the players alternate but $L(\mathcal C_1)$ does not contain a vertex of the environment). The teacher replies with an existential implication counterexample, say $(s \vbar \vbar, \mathcal A)$ with $L(\mathcal A) = \{e \vbar \vbar, e \vbar \vbar \vbar\}$.

In the next round, let us assume that the learner conjectures the \dfa $\mathcal C_2$ with $L(\mathcal C_2) = \{ s \vbar^n \mid n \geq 2 \} \cup \{ e \vbar^m \mid m \geq 3 \}$. This conjecture passes all checks (i.e., $L(\mathcal C_2)$ is a winning set), the teacher replies ``yes’’, and the learning ends. \exend
\end{example}

It is important to note that classical learning frameworks for regular languages that involve learning from positive and negative data only, such as Gold’s passive learning~\cite{DBLP:journals/iandc/Gold78} or Angluin’s active learning~\cite{DBLP:journals/iandc/Angluin87}, are insufficient in our setting. If the learner provides a conjecture $\mathcal C$ that violates Condition~\ref{main:cond:winning_set:3} or \ref{main:cond:winning_set:4} of Definition~\ref{main:def:winning_set}, the teacher is stuck. For instance, if $\mathcal C$ does not satisfy  Condition~\ref{main:cond:winning_set:4}, the teacher does not know whether to exclude $u$ or to include $E(\{u\})$. Returning an implication counterexample, however, resolves this problem in that it communicates exactly why the conjecture is incorrect and, hence, allows the learner to make progress.\kern-.06em\footnote{Garg et.~al.~\cite{DBLP:conf/cav/0001LMN14} argue comprehensively why implications needed in a robust invariant learning framework. Their arguments also apply to our setting as one obtains a setting similar to Garg et.~al.’s by considering a solitary game with \playerb as the only player.}


%


\section{A Generic Teacher}
\label{sec:teacher}
We now present a generic teacher that, taking a rational safety game as input, answers queries according to Definition~\ref{main:def:teacher}.
For the remainder of this section, fix a rational safety game $\mathfrak G_\Sigma = (\mathfrak A_\Sigma, \mathcal A_F, \mathcal A_I)$ over the rational arena $\mathfrak A_\Sigma = (\mathcal A_{V_0}, \mathcal A_{V_1}, \mathcal T_E)$, and let $\mathcal C$ be a \dfa conjectured by the learner.

To answer a query, the teacher performs Checks~\ref{main:check:teacher:1} to \ref{main:check:teacher:4} of Definition~\ref{main:def:teacher} as described below. If the conjecture passes all checks, the teacher returns ``yes’’; otherwise, he returns a corresponding counterexample, as described next.

\paragraph*{Check~\ref{main:check:teacher:1} (initial vertices)}
The teacher computes an \nfa $\mathcal B$ with $L(\mathcal B) = L(\mathcal A_I) \setminus L(\mathcal C)$. If $L(\mathcal B) \neq \emptyset$, he returns a positive counterexample $u \in L(\mathcal B)$.

\paragraph*{Check~\ref{main:check:teacher:2} (safe vertices)}
The teacher computes an \nfa $\mathcal B$ with $L(\mathcal B) = L(\mathcal C) \setminus  L(\mathcal A_F)$. If $L(\mathcal B) \neq \emptyset$, he returns a negative counterexample $u \in L(\mathcal B)$.

\paragraph*{Check~\ref{main:check:teacher:3} (existential closure)}
To check existential closure, the teacher successively computes three \nfas:
\begin{enumerate}
	\item An \nfa $\mathcal B_1$ with $L(\mathcal B_1) = R(\mathcal T_E)^{-1}(L(\mathcal C))$; the language $L(\mathcal B_1)$ contains all vertices that have a successor in $L(\mathcal C)$.
	\item An \nfa $\mathcal B_2$ with $L(\mathcal B_2) = L(\mathcal A_{V_0}) \setminus L(\mathcal B_1)$; the language $L(\mathcal B_2)$ contains all vertices of \playera that have no successor in $L(\mathcal C)$.
	\item An \nfa $\mathcal B_3$ with $L(\mathcal B_3) = L(\mathcal C) \cap L(\mathcal B_2)$; the language $L(\mathcal B_3)$ contains all vertices of \playera that belong to $L(\mathcal C)$ and have no successor in $L(\mathcal C)$.
\end{enumerate}
Every $u \in L(\mathcal B_3)$ is a witness that $\mathcal C$ is not existentially closed. Hence, if $L(\mathcal B_3) \neq \emptyset$, the teacher picks an arbitrary $u \in L(\mathcal B_3)$ and returns the existential implication counterexample $(u, \mathcal A)$ where $L(\mathcal A) = R(\mathcal T_E)(\{u\})$.

\paragraph*{Check~\ref{main:check:teacher:4} (universal closure)}
To check universal closure, the teacher, again, computes three \nfas:
\begin{enumerate}
	\item An \nfa $\mathcal B_1$ with $L(\mathcal B_1) = \bigl( L(\mathcal A_{V_0}) \cup L(\mathcal A_{V_1}) \bigr) \setminus L(\mathcal C)$; the language $L(\mathcal B_1)$ contains all vertices not in $L(\mathcal C)$.
    \item An \nfa $\mathcal B_2$ with $L(\mathcal B_2) = R(\mathcal T_E)^{-1}(L(\mathcal B_1))$; the language $L(\mathcal B_2)$ contains all vertices that have a successor not belonging to $L(\mathcal C)$.
    \item An \nfa $\mathcal B_3$ with $L(\mathcal B_3) = L(\mathcal A_{V_1}) \cap L(\mathcal C) \cap L(\mathcal B_2)$; the language $L(\mathcal B_3)$ contains all vertices of \playerb that are in $L(\mathcal C)$ and have at least one successor not in $L(\mathcal C)$.
\end{enumerate}
Every $u \in L(\mathcal B_3)$ is a witness that $\mathcal C$ is not universally closed. Hence, if $L(\mathcal B_3) \neq \emptyset$, the teacher picks an arbitrary $u \in L(\mathcal B_3)$ and returns the universal implication counterexample $(u, \mathcal A)$ where $L(\mathcal A) = R(\mathcal T_E)(\{u\})$.

All checks can be performed using standard methods of automata theory, including product constructions, projections, determinizing automata, and emptiness checks (see Lemma~\ref{main:lem:transducer}).


\section{A Learner for Rational Safety Games}
\label{sec:learner}

We design our learner with two key features:
\begin{enumerate*}[label=(\arabic*)]
	\item the learner always conjectures a \dfa consistent with the counterexamples received so far (we make this precise shortly), and
	\item the learner always conjectures a minimal consistent \dfa (i.e., a \dfa with the least number of states among all \dfas that are consistent with the received counterexamples).
\end{enumerate*}
The first design goal prevents the learner from making the same mistake twice, while the second design goal facilitates convergence of the overall learning (assuming that a winning set exists).

To meet these goals, our learner stores counterexamples in a data structure, which we call \emph{sample}. Formally, a \emph{sample} is a four-tuple $\mathcal S = (\Po, \Ne, \Ex, \Un)$ consisting of a finite set $\Po \subset \Sigma^\ast$ of positive words, a finite set $\Ne \subset \Sigma^\ast$ of negative words, a finite set $\Ex \subset \Sigma^\ast \times \mathit{NFA}_\Sigma$ of existential implications, and a finite set $\Un \subset \Sigma^\ast \times \mathit{NFA}_\Sigma$ of universal implications. We encourage the reader to think of a sample as a finite approximation of the safety game learned thus far.

In every iteration, our learner constructs a minimal \dfa \emph{consistent} with the current sample. A \dfa $\mathcal B$ is called \emph{consistent} with a sample $\mathcal S = (\Po, \Ne, \Ex, \Un)$ if
\begin{enumerate}
	\item \label{main:cond:consistent_positive} $\Po \subseteq L(\mathcal B)$;
	\item \label{main:cond:consistent_negative} $\Ne \cap L(\mathcal B) = \emptyset$;
	\item \label{main:cond:consistent_existential} $u \in L(\mathcal B)$ implies $L(\mathcal B) \cap L(\mathcal A) \neq \emptyset$ for each $(u, \mathcal A) \in \Ex$;
	\item \label{main:cond:consistent_universal} $u \in L(\mathcal B)$ implies $L(\mathcal A) \subseteq L(\mathcal B)$ for each $(u, \mathcal A) \in \Un$.
\end{enumerate}

Constructing a \dfa that is consistent with a sample is possible only if the sample does not contain contradictory information.
Contradictions can arise in two ways: first, $\Po$ and $\Ne$ are not disjoint; second, the (alternating) transitive closure of the implications in $\Ex$ and $\Un$ contains a pair $(u, v)$ with $u \in \Po$ and $v \in \Ne$.
This observation justifies to introduce the notion of \emph{contradiction-free} samples: a sample $\mathcal S$ is called \emph{contradiction-free} if a \dfa that is consistent with $\mathcal S$ exists. Since we assume that \playera wins from set $I$, a winning set exists and the counterexamples returned by the teacher always form contradiction-free samples.\kern-.06em\footnote{In fact, checking for contradictions equips the learner with a means to detect that the game is won by \playerb. However, since determining the winner of a rational safety game is undecidable, any sample obtained during the learning might be contradiction-free despite the fact that \playerb wins.}

After having constructed  a minimal consistent \dfa, the learner conjectures it to the teacher. If the teacher replies ``yes'', the learning terminates. If the teacher returns a counterexample, on the other hand, the learner adds it to the appropriate set in $\mathcal S$ 
and iterates. This procedure is sketched as Algorithm~\ref{main:alg:sat_learner}. Note that, by definition of the teacher, a conjecture is guaranteed to accept a wining set once the learning terminates.

\begin{algorithm}[t]
	Initialize an empty sample $\mathcal S = (\Po, \Ne, \Ex, \Un)$ with $\Po = \emptyset$, $\Ne = \emptyset$, $\Ex = \emptyset$, and $\Un = \emptyset$\;

	\Repeat{the teacher replies ``yes'' to an equivalence query}
	{
		Construct a minimal \dfa $\mathcal A_\mathcal S$ consistent with $\mathcal S$\;
		Submit $\mathcal A_\mathcal S$ to an equivalence query\;
		\If{the teacher returns a counterexample}
		{
			Add the counterexample to $\mathcal S$\;
		}
	}
	
	\Return $\mathcal A_\mathcal S$\;

	\caption{A learner for rational safety games}\label{main:alg:sat_learner}
\end{algorithm}

It is left to describe how the learner actually constructs a minimal \dfa that is consistent with the current sample. However, this task, known as \emph{passive learning}, is computationally hard (i.e., the corresponding decision problem is NP-complete) already in the absence of implications \cite{DBLP:journals/iandc/Gold78}.
Our strategy to approach this hurdle is to translate the original problem into a sequence of satisfiability problems of formulas in propositional Boolean logic and use highly optimized constraint solvers as a practically effective means to solve the resulting formulas (note that a translation into a logical formulation is a popular and effective strategy).
More precisely, our learner creates and solves propositional Boolean formulas $\varphi_n^\mathcal S$, for increasing values of $n \in \mathbb N$, $n \geq 1$, with the following two properties:
\begin{enumerate}
	\item \label{main:prop:sat_learner_1} The formula $\varphi_n^\mathcal S$ is satisfiable if and only if there exists a \dfa with $n$ states that is consistent with $\mathcal S$.
	\item \label{main:prop:sat_learner_2} A model $\mathfrak M$ of $\varphi_n^\mathcal S$ (i.e., a satisfying assignment of the variables in $\varphi_n^\mathcal S$) contains sufficient information to construct a \dfa, denoted by $\mathcal A_\mathfrak M$, that has $n$ states and is consistent with $\mathcal S$.
\end{enumerate}

If $\varphi_n^\mathcal S$ is satisfiable, then Property~\ref{main:prop:sat_learner_2} enables us to construct a consistent \dfa from a model. However, if the formula is unsatisfiable, then the parameter $n$ has been chosen too small and the learner increments it (e.g., by one or using a binary search). This procedure is summarized as Algorithm~\ref{main:alg:consistent_dfa}. We show its correctness shortly in Section~\ref{sec:correctness_sat_learner}.

\begin{algorithm}[t]
	\BlankLine
	\KwIn{A contradiction-free sample $\mathcal S$}
	\KwOut{A minimal \dfa that is consistent with $\mathcal S$}

	$n \gets 0$\;
	\Repeat{$\varphi_n^\mathcal S$ is satisfiable, say with model $\mathfrak M$}{
		$n \gets n+1$\;
		Construct and solve $\varphi_n^\mathcal S$\;
	}
	\Return $\mathcal A_\mathfrak M$\;

	\caption{Computing a minimal consistent \dfa.}\label{main:alg:consistent_dfa} 
\end{algorithm}

The key idea of the formula $\varphi_n^\mathcal S$ is to encode a \dfa with $n$ states by means of Boolean variables and to pose constraints on those variables. Our encoding relies on a simple observation:
for every \dfa there exists an isomorphic (hence, equivalent) \dfa over the state set $Q = \{ 0, \ldots, n-1 \}$ with initial state $q_0 = 0$; moreover, given that $Q$ and $q_0$ are fixed, any \dfa with $n$ states is uniquely determined by its transitions and final states. Therefore, we can fix the state set of the prospective \dfa as $Q = \{ 0,  \ldots, n-1 \}$ and the initial state as $q_0 = 0$; the alphabet $\Sigma$ is announced by the teacher.

Our encoding of transitions and final states follows an idea from~\cite{DBLP:conf/nfm/NeiderJ13} (independently due to~\cite{DBLP:conf/icgi/HeuleV10}). We introduce Boolean variables $d_{p, a, q}$ and $f_q$ where $p, q \in Q$ and $a \in \Sigma$, which have the following meaning: setting $d_{p, a, q}$ to \true means that the transition $\delta(p, a) = q$ exists in the prospective \dfa, and setting $f_q$ to \true means that $q$ is a final state.

To make sure that the variables $d_{p, a, q}$ encode a deterministic transition function, we impose two constraints:
\begin{align}
	\bigwedge_{p \in Q}~ \bigwedge_{a \in \Sigma}~ \bigwedge_{q, q' \in Q, q \neq q'}~ \lnot d_{p, a, q} \lor \lnot d_{p, a, q'} \label{main:for:dfa_1} \\
	\bigwedge_{p \in Q}~ \bigwedge_{a \in \Sigma}~ \bigvee_{q \in Q}~ d_{p, a, q} \label{main:for:dfa_2}
\end{align}


Let $\varphi_n^\text{DFA}$ be the conjunction of Formulas~\eqref{main:for:dfa_1} and \eqref{main:for:dfa_2}. Given a model $\mathfrak M$ of $\varphi_n^\text{DFA}$ (we assume a model to be a map from the variables of a formula to the set $\{true, false\}$), deriving the encoded \dfa is straightforward, as shown next.

\begin{definition}\label{main:def:conjecture_from_SAT}
\itshape
Let $\mathfrak M$ be a model of $\varphi_n^\text{DFA}$. We define the \dfa $\mathcal A_\mathfrak M = (Q, \Sigma, q_0, \delta, F)$ by
\begin{enumerate*}[label=(\arabic*), ref=\arabic*]
    \item $\delta(p, a) = q$ for the unique $q \in Q$ with $\mathfrak M(d_{p, a, q}) = \true$; and 
	\item $F = \{ q \in Q \mid \mathfrak M(f_q) = \true \}$.
\end{enumerate*}
(Recall that we fixed $Q = \{0, \ldots, n-1 \}$ and $q_0 = 0$.) \defend
\end{definition}

To enforce that $\mathcal A_\mathfrak M$ is consistent with the given sample $\mathcal S = (\Po, \Ne, \Ex, \Un)$, we impose further constraints, corresponding to the four requirements of consistent \dfas:
\begin{itemize}
	\item a formula $\varphi_n^\Po$ asserting $\Po \subseteq L(\mathcal A_\mathfrak M)$;
	\item a formula $\varphi_n^\Ne$ asserting $\Ne \cap L(\mathcal A_\mathfrak M) = \emptyset$;
	\item a formula $\varphi_n^\Ex$ asserting that $u \in L(\mathcal A_\mathfrak M)$ implies $L(\mathcal A_\mathfrak M) \cap L(\mathcal A) \neq \emptyset$ for each $(u, A) \in \Ex$; and
    \item a formula $\varphi_n^\Un$ asserting that $u \in L(\mathcal A_\mathfrak M)$ implies $L(\mathcal A_\mathfrak M) \subseteq L(\mathcal A)$ for each $(u, A) \in \Un$.
\end{itemize}
Then,
$\varphi_n^\mathcal S \coloneqq \varphi_n^\text{DFA} \land \varphi_n^\Po \land \varphi_n^\Ne \land \varphi_n^\Ex \land \varphi_n^\Un$.
We here sketch formula $\varphi_n^\Un$ and refer the reader to Appendix~\ref{apdx:formula_phi} for a detailed presentation of the remaining formulas. A description of $\varphi_n^\Po$ and $\varphi_n^\Ne$ can also be found in~\cite{DBLP:conf/nfm/NeiderJ13}.

\subsection{\texorpdfstring{The formula $\varphi_n^\Un$}{The formula phi}}
We break the construction of $\varphi_n^\Un$ down into smaller parts. Roughly speaking, we construct for each universal implication $\iota = (u, \mathcal A) \in \Un$ a formula $\varphi_n^\iota$ that asserts $L(\mathcal A)\subseteq L(\mathcal A_\mathfrak M)$ if $u \in L(\mathcal A_\mathfrak M)$. The formulas $\varphi_n^\Un$ is then the finite conjunction $\bigwedge_{\iota \in \Un} \varphi_n^\iota$. For the remainder, let us fix a universal implication $\iota \in \Un$, say $\iota = (u, A)$ with $A = (Q_\mathcal A, \Sigma, q_0^\mathcal A, \Delta_\mathcal A, F_\mathcal A)$, and let $\ante(\Un) = \{ u \mid (u, \mathcal A) \in \Un \}$ be the set of all words occurring as antecedent of a universal implication.

As a preparatory step, we introduce auxiliary Boolean variables that track the runs of $\mathcal A_\mathfrak M$ on words of $\pref(\ante(\Un))$ in order to detect when $\mathcal A_\mathfrak M$ accepts the antecedent of a universal implication. More precisely, we introduce variables $x_{u, q}$ where $u \in \pref(\ante(\Un))$ and $q \in Q$, which have the meaning that $x_{u, q}$ is set to \true if $\mathcal A_\mathfrak M \colon q_0 \myrightarrow{u} q$ (i.e., $\mathcal A_\mathfrak M$ reaches state $q$ on reading $u$):
\vspace{-\baselineskip}
\begin{align}
	x_{\varepsilon, q_0} \label{main:for:run_1} \\
    \bigwedge_{u \in \pref(\ante(\Un))}~ \bigwedge_{q \neq q' \in Q}~ \neg x_{u, q} \vee \neg x_{u,q'} \label{main:for:run_2} \\
    \bigwedge_{ua \in \pref(\ante(\Un))}~ \bigwedge_{p,q \in Q}~ (x_{u, p} \wedge d_{p, a, q}) \rightarrow x_{ua, q} \label{main:for:run_3}
\end{align}
Formula~\eqref{main:for:run_1} asserts that $x_{\varepsilon, q_0}$ is set to \true since any run starts in the initial state $q_0$. Formula~\eqref{main:for:run_2} enforces that for every $u \in \pref(\ante(\Un))$ there exists at most one $q \in Q$ such that $x_{u, q}$ is set to \true (in fact, the conjuction of Formulas~\eqref{main:for:dfa_2}--\eqref{main:for:run_3} implies that there exists a unique such state). Finally, Formula~\eqref{main:for:run_3} prescribes how the run of $\mathcal A_\mathfrak M$ on a word $u \in \pref(\ante(\Un))$ proceeds: if $\mathcal A_\mathfrak M$ reaches state $p$ on reading $u$ (i.e., $x_{u, p}$ is set to $\true$) and there exists a transition from $p$ to state $q$ on reading the symbol $a \in \Sigma$ (i.e., $d_{p, a, q}$ is set to \true), then $\mathcal A_\mathfrak M$ reaches state $q$ on reading $ua$ and $x_{ua}$ needs to be set to \true.

We now define $\varphi_n^\iota$. The formula ranges, in addition to $d_{p, a, q}$, $f_q$, and $x_{u, q}$, over Boolean variables $y_{q, q'}^\iota$ where $q \in Q$ and $q' \in Q_\mathcal A$, which track runs of $\mathcal A$ and $\mathcal A_\mathfrak M$. Their precise meaning is the following: if there exists a word $u \in \Sigma^\ast$ with $\mathcal A_\mathfrak M \colon q_0 \myrightarrow{u} q$ and $\mathcal A \colon q_0^\mathcal A \myrightarrow{u} q'$, then $y_{q, q'}^\iota$ is set to \true:
\vspace{-.25\baselineskip}
\begin{align}
	y_{q_0, q_0^\mathcal A}^\iota \label{main:for:universal_1} \\
	\bigwedge_{p, q \in Q}~ \bigwedge_{(p', a, q') \in \Delta_\mathcal A}~ (y_{p, p'}^\iota \land d_{p, a, q}) \rightarrow y_{q, q'}^\iota \label{main:for:universal_2}
\end{align}
Formula~\eqref{main:for:universal_1} enforces $y_{q_0, q_0^\mathcal A}^\iota$ to be set to \true because $\mathcal A_\mathfrak M \colon q_0 \myrightarrow{\varepsilon} q_0$ and $\mathcal A \colon q_0^\mathcal A \myrightarrow{\varepsilon} q_0^\mathcal A$. Formula~\eqref{main:for:universal_2} is similar to Formula~\eqref{main:for:run_3} and describes how the runs of $\mathcal A_\mathfrak M$ and $\mathcal A$ proceed: if there exists a word $v$ such that $\mathcal A_\mathfrak M \colon q_0 \myrightarrow{v} p$ and $\mathcal A \colon q_0^\mathcal A \myrightarrow{v} p'$ (i.e., $y_{p, p'}^\iota$ is set to \true) and there are transitions $(p', a, q') \in \Delta_\mathcal A$ and $\delta(p, a) = q$ in $\mathcal A_\mathfrak M$, then $\mathcal A_\mathfrak M \colon q_0 \myrightarrow{va} q$ and $\mathcal A \colon q_0^\mathcal A \myrightarrow{va} q'$, which requires $y_{q, q'}^\iota$ to be set to \true. 

Finally, the next constraint ensures that whenever $\mathcal A_\mathfrak M$ accepts $u$ (i.e., the antecedent is \true), then all words that lead to an accepting state in $\mathcal A$ also lead to an accepting state in $\mathcal A_\mathfrak M$ (i.e., the consequent is \true).
\begin{align}
	\bigl( \bigvee_{q \in Q}~ x_{u, q} \land f_q \bigr) \rightarrow \bigl( \bigwedge_{q \in Q}~ \bigwedge_{q' \in F_\mathcal A}~ y_{q, q'}^\iota \rightarrow f_q \bigr) \label{main:for:universal_3}
\end{align}

Let $\varphi_n^{\ante(\Un)}$ be the conjunction of Formulas~\eqref{main:for:run_1}, \eqref{main:for:run_2}, and \eqref{main:for:run_3} as well as $\varphi_n^\iota$ the conjunction of Formulas~\eqref{main:for:universal_1}, \eqref{main:for:universal_2}, and \eqref{main:for:universal_3}. Then, $\varphi_n^\Un$ is the (finite) conjunction $\varphi_n^{\ante(\Un)} \land \bigwedge_{\iota \in \Un} \varphi_n^\iota$.

\subsection{Correctness of the Learner}\label{sec:correctness_sat_learner}
We now sketch a correctness proof of the learner---we refer the reader to Appendix~\ref{apdx:sat_learner_proof} for a detailed proof. First, we state that $\varphi_n^\mathcal S$ has the desired properties.

\begin{lemma}\label{main:lem:phi_correct}
\itshape
Let $\mathcal S$ be a sample, $n \geq 1$, and 
$ \varphi_n^\mathcal S$ be as defined above.
Then, the following statements hold:
\begin{enumerate*}[label=(\arabic*), ref=\arabic*]
	\item \label{main:stmt:phi_correct:1} If $\mathfrak M \models \varphi_n^\mathcal S$, then $\mathcal A_\mathfrak M$ is a \dfa with $n$ states that is consistent with $\mathcal S$.
	\item \label{main:stmt:phi_correct:2} If  there exists a \dfa that has $n$ states and is consistent with $\mathcal S$, then $\varphi_n^\mathcal S$ is satisfiable.
\end{enumerate*}
\end{lemma}

Next, let us show the correctness of Algorithm~\ref{main:alg:consistent_dfa}. 

\begin{theorem}\label{main:thm:consistent_dfa_learner_correct}
\itshape
Given a contradiction free-sample $\mathcal S$, Algorithm~\ref{main:alg:consistent_dfa} returns a minimal \dfa (in terms of the number of states) that is consistent with $\mathcal S$. If a minimal consistent \dfa has $k$ states, then Algorithm~\ref{main:alg:consistent_dfa} terminates after $k$ iterations.
\end{theorem}

\begin{IEEEproof}
Given a sample $\mathcal S$, suppose that there exists a \dfa that has $k$ states and is consistent with $\mathcal S$. Then, $\varphi_n^\mathcal S$ is satisfiable for all $n \geq k$ (see Lemma~\ref{main:lem:phi_correct}). Moreover, if $\mathfrak M$ is a model of $\varphi_n^\mathcal S$, then $\mathcal A_\mathfrak M$ is a \dfa with $n$ states that is consistent with $\mathcal S$. Since Algorithm~\ref{main:alg:consistent_dfa} increases the parameter $n$ by one in every iteration (starting with $n=1$), the algorithm eventually finds the smallest value for which $\varphi_n^\mathcal S$ is satisfiable (after $k$ iterations) and, hence, a consistent \dfa of minimal size.
\end{IEEEproof}

Finally, we can prove the correctness of our learner.

\begin{theorem}\label{main:thm:sat_learner_correct}
\itshape
Given a teacher, Algorithm~\ref{main:alg:sat_learner}, equipped with Algorithm~\ref{main:alg:consistent_dfa} to construct conjectures, terminates and returns a (minimal) \dfa accepting a winning set if one exists.
\end{theorem}

%

\begin{IEEEproof}
Theorem~\ref{main:thm:sat_learner_correct} follows from three observations about the learner:
\begin{enumerate*}[label=(\arabic*), ref=\arabic*]
	\item \label{main:obs:correctness_learner:1} The learner never conjectures the same \dfa twice (due to Theorem~\ref{main:thm:consistent_dfa_learner_correct} and the fact that counterexamples are added to the sample).
	\item \label{main:obs:correctness_learner:2} The conjectures grow monotonically in size (due to minimality of conjectures) with increasing $n$, and
	\item \label{main:obs:correctness_learner:3} adding counterexamples to a sample does not rule out any solution (as every \dfa accepting a winning set is consistent with any sample produced during the learning).
\end{enumerate*}
Now, suppose a \dfa accepting a winning set exists, say with $k$ states. Due to Observations~\ref{main:obs:correctness_learner:1} and \ref{main:obs:correctness_learner:2}, the learner eventually conjectures a \dfa with $k$ states and, moreover, cannot conjecture a larger \dfa (due to Observation~\ref{main:obs:correctness_learner:3} and the minimality of conjectures). Hence, the learner eventually conjectures a \dfa with $k$ states that accepts a winning set, and the learning terminates.
\end{IEEEproof}


\section{Experiments}
\label{sec:experiments}

In order to demonstrate the feasibility of our learning approach, we implemented a Java prototype  using the \textsc{Brics} automaton library~\cite{brics} and Microsoft's Z3~\cite{DBLP:conf/tacas/MouraB08} constraint solver. The source code, including the games used in the experiments, is available at \url{http://preview.tinyurl.com/n7a7byj}.

In addition to the learner of Section~\ref{sec:learner}, we implemented a learner based on the popular RPNI algorithm~\cite{OncinaGarcia92}, which is a polynomial time algorithm for learning \dfas from positive and negative words. For this learner, we modified the RPNI algorithm such that it constructs a consistent \dfa from existential and universal implications in addition to positive and negative words (a detailed presentation can be found in Appendix~\ref{apdx:rpni_learner}). In contrast to Algorithm~\ref{main:alg:consistent_dfa}, our modified version of RPNI cannot guarantee to find smallest consistent \dfas and, hence, the resulting learner is a fast heuristic that is sound but in general not complete. Another limitation is that it can only handle implication counterexamples of the form $(u, \mathcal A)$ where $L(\mathcal A)$ is finite.
We refer to the learner of Section~\ref{sec:learner} as \emph{SAT learner} and the RPNI-based learner as \emph{RPNI learner}.

\begin{table*}[t]
	\centering
	\caption{Results of the first benchmark suite}\label{main:tbl:various_examples}
	\begin{tabular}{@{\hskip -.15em}lr@{\hskip 1.75em}*{7}{r}@{\hskip 1.75em}*{7}{r}}
		\toprule
		& & \multicolumn{7}{c}{SAT learner} & \multicolumn{7}{c}{RPNI learner} \\
		\cmidrule(r{1.75em}){3-9} \cmidrule{10-16}
		Game & $|\mathfrak G|$ & Time in s & Iter. & Size & $|\Po|$ & $|\Ne|$ & $|\Ex|$ & $|\Un|$ & Time in s & Iter. &  Size & $|\Po|$ & $|\Ne|$ & $|\Ex|$ & $|\Un|$ \\
		\midrule
		Diagonal & 29 & 1.352 & 62 & 4 & 1 & 55 & 2 & 3 & 1.000 & 77 & 6 & 1 & 54 & 10 & 11 \\
		Box & 25 & 0.516 & 32 & 4 & 1 & 30& 0 & 0 & 0.188 & 15 & 5 & 1 & 10 & 1 & 2 \\
		Solitary Box & 22 & 4.289 & 81 & 6 & 1 & 77 & 2 & 0 & 0.156 & 16 & 6 & 1 & 13 & 1 & 0 \\
		Follow & 53 & 165.670 & 294 & 7 & 2 & 269 & 10 & 12 & \multicolumn{7}{c}{\leaders\hrule depth -2pt height 2.4pt\hfill timeout ($>$ 300\,s)\leaders\hrule depth -2pt height 2.4pt\hfill \mbox{}} \\
		Evasion & 56 & 140.888 & 255 & 7 & 2 & 232 & 11 & 9 & 2.316 & 142 & 12 & 1 & 115 & 14 & 11 \\
		Program-repair & 41 & 1.948 & 62 & 3 & 2 & 55 & 4 & 0 & 0.438 & 31 & 4 & 1 & 20 & 9 & 0 \\
		\bottomrule
	\end{tabular}
	\vspace{-1\baselineskip}
\end{table*}

Our experiments are on a slightly restricted type of games:
\begin{enumerate}
	\item \label{main:itm:exp:restriction:1} Edge relations are \emph{automatic}. Automatic relations are defined by transducers that do not possess transitions of the form $(a, \varepsilon)$ and $(\varepsilon, b)$ but rather use a dedicated padding symbol to balance the length of their input-words.\kern-.06em\footnote{Automatic relations constitute a proper subset of rational relations, but are still expressive enough to encode computations of Turing machines.}
	\item \label{main:itm:exp:restriction:2} Each vertex of an arena has a finite (but not necessarily bounded) number of outgoing edges.
\end{enumerate}
Restriction~\ref{main:itm:exp:restriction:1} simplifies the implementation of the teacher. Restriction~\ref{main:itm:exp:restriction:2} is due to the limitation of the RPNI learner.

We use two benchmark suits: the first suite serves to demonstrate the feasibility of our techniques for various examples, predominantly taken from the area of motion planning; the second suite serves to assess the performance of our techniques when confronted with games of increasing ``complexity''. All games were given as finite automata, and we employed the teacher described in Section~\ref{sec:teacher}. We conducted all experiments on an Intel Core i7-4510U CPU (running Microsoft Windows 8.1) with a memory limit of $4$\,GiB and a runtime limit of $300$\,s. 

\subsection{Examples}
We consider the following examples. 
\begin{description}[font={\normalfont\itshape}, align=left, leftmargin=*]
	\item[Diagonal game:] A robot moves on an infinite, discrete two-dimensional grid world from one cell to an adjacent cell. \playera controls the robot's vertical movement, whereas \playerb controls the horizontal movement. Both players move the robot in alternation, and \playera's objective is to stay inside a margin of two cells around the diagonal.
	\item[Box game:] A version of the diagonal game in which \playera's objective is to stay within a horizontal stripe of width three.  
	\item[Solitary box game:] A version of the box game in which \playera is the only player and has control over both the horizontal and the vertical movement.
	\item[Evasion game:] Two robots move in alternation on an infinite, two-dimensional grid. Each robot is controlled by a player. \playera's objective is to avoid collision with \playerb's robot.
	\item[Follow game:] A version of the evasion game in which \playera's objective is to keep his robot within a distance of two cells (in the Manhattan distance) from \playerb's robot.
	\item[Program-repair game:] A finitely-branching version of the program-repair game described by Beyene et~al.~\cite{DBLP:conf/popl/BeyeneCPR14}.
\end{description}

Table~\ref{main:tbl:various_examples} lists the overall time taken by each of the two learners to learn a winning set (including the time taken by the teacher) as well as further statistics of the learning process. The second column $|\mathfrak G|$ corresponds to sum of states of all automata constituting a game (\emph{size of the game}), which serves as measure for the complexity of a game. The remaining columns list the number of iterations, the number of states of the learned \dfa, and the cardinality of each set of the final sample.

As Table~\ref{main:tbl:various_examples} shows, the SAT learner computed the winning sets for all games, whereas the RPNI learner computed the winning sets for all but the Follow game. Since the RPNI learner does not compute minimal consistent \dfas, we expected that it is on average faster than the SAT learner, which turned out to be the case. However, the RPNI learner fails to terminate within the time limit on the Follow game, and the large number of iterations seem to indicate that the learner in fact diverges.
  
Finally, it is important to note that the teacher replied implication counterexamples in all but one experiment. This observation highlights that classical learning algorithms, which learn from positive and negative words only, are insufficient to learn winning sets (since the learning would be stuck at that point) and one has to move to a richer learning framework.

\subsection{Scalability Benchmarks}

To assess the scalability of our technique when confronted with inputs of increasing size, we modified the game of Example~\ref{main:ex:rational_safety_game} such that the safe region is now determined by two parameters, namely $k$ and \parmmax, and contains all positions in the interval $[k, \parmmax]$ (we assume $k < \parmmax$ and fix $k = 1$). In this new setting, the number of states of the automaton $\mathcal A_F$ increases when \parmmax increases as the automaton needs to count in unary to check the position of the robot.

Figure~\ref{main:fig:scalability} depicts the overall time taken to learn a winning set, depending on the parameter \parmmax. To put the runtimes into perspective, it also shows the size of the games.

\pgfplotsset{scalability/.style={
	shorten >=0pt,
	width=.85\linewidth,
	height=4.4cm,
	every axis x label/.append style={yshift=.3\baselineskip},
	font=\small,
	tick align=inside,
	xtick pos=left,
	log origin=0,
	legend style={draw=none, font=\footnotesize, legend pos=north west, cells={anchor=west}},
	mark options={draw=black, fill=black!10, thin, solid},
}}

\begin{figure}
	\centering

	\begin{tikzpicture}
	\begin{loglogaxis}[
			scalability,
			axis y line*=left,
			ymax=1000,
			ymin=.1,
			xtick={0, 10, 100, 1000, 10000, 100000},
			ytick={0, .1, 1, 10, 100, 1000},
			xlabel={\parmmax},
			ylabel={time in s}]
			
			\addplot[x={max}, y={total time}, mark=*, densely dotted] table[col sep=comma] {scalability_sat.csv}; \label{main:sat_plot}
			\addlegendentry{SAT learner}
			\addplot[x={max}, y={total time}, mark=square*, densely dotted] table[col sep=comma] {scalability_rpni.csv}; \label{main:rpni_plot}
			\addlegendentry{RPNI learner}
		\end{loglogaxis}
		
		\begin{loglogaxis}[
			scalability,
			axis y line*=right,
			axis x line=none,
			ymin=1,
			ylabel={$|\mathfrak G|$}]
			
			\addlegendimage{/pgfplots/refstyle=main:sat_plot}\addlegendentry{SAT learner}
			\addlegendimage{/pgfplots/refstyle=main:rpni_plot}\addlegendentry{RPNI learner}
			
			\addplot[x={max}, y={size}, mark=x] table[col sep=comma] {scalability_size.csv};
			\addlegendentry{$|\mathfrak G|$}
		\end{loglogaxis}
	\end{tikzpicture}
	\vskip -.75\baselineskip
	\caption{Results of the scalability benchmark.\label{main:fig:scalability}} \vspace{-0.1in}
	\vskip .25\baselineskip
\end{figure}
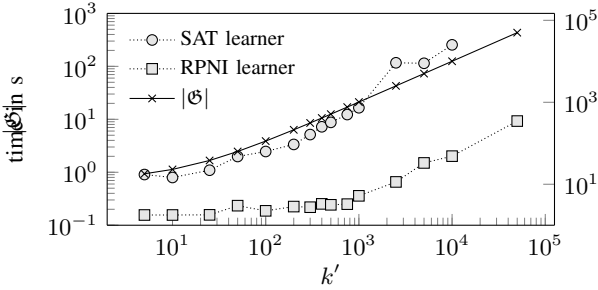

On the scalability benchmark suite, the RPNI learner was about one order of magnitude faster than the  SAT learner and can computed a winning set for games up to a combined size of $50\,000$. The SAT learner, on the other hand, computed a winning set for games up to a combined size of $10\,000$ but did not terminate for game with $\parmmax = 50\,000$. While a thorough assessment remains as part of future work, our results promise applicability to practically interesting problem instances.


\section{Conclusion}
We developed an automata learning method to construct finite-state reactive
controllers for systems whose interactions with their environment are modeled by infinite-state  games. We focused on the practically interesting family of safety games, utilized a symbolic representation of the underlying game, developed specific implementations of the learner and the teacher, and demonstrated the feasibility of the method on a set of problems motivated by robotic motion planning.


\bibliographystyle{IEEEtran}
\bibliography{IEEEabrv,references}

\cleardoublepage
\appendices

\newcommand{\false}{\ensuremath{\mathit{false}}\xspace}
\newcommand{\words}{\ensuremath{W}}
\newcommand{\mysection}[1]{\subsection*{\normalfont \scshape \leaders\hrule depth -2pt height 2.4pt\hfill #1 \leaders\hrule depth -2pt height 2.4pt\hfill \mbox{}}}

\section{\texorpdfstring{Constructing Consistent \dfas \\ Using Constraint Solvers}{Constructing Consistent \dfas Using Constraint Solvers}}
\label{apdx:formula_phi}
The key building block of our learner is an algorithm that, given a sample $\mathcal S$, produces a smallest \dfa that is consistent with $\mathcal S$. Recall that the learner translates this problem into a series of satisfiability problem of propositional Boolean formulas $\varphi_n^\mathcal S$ and uses a constraint solver to check their satisfiability. 

In the following, we describe in detail how the formula $\varphi_n^\mathcal S$ is constructed. For the sake of a self-contained presentation, we repeat parts of Section~\ref{sec:learner}; as a beneficial side-effect, this repetition allows us to provide further explanations of the formulas presented in Section~\ref{sec:learner}. Moreover, to facilitate a more concise and accessible description, we define $\varphi_n^\mathcal S$ slightly different. In particular, we introduce a formula $\varphi^\words_n$, which tracks the run of $\mathcal A_\mathfrak M$ on words occurring in the sample (in $\Po$, $\Ne$, and as antecedent of an implication). In contrast to Section~\ref{sec:learner} (where we defined the formula $\varphi_n^\Un$ to track the run of $\mathcal A_\mathfrak M$ on the set $\ante(\Un)$) this approach results in more concise and easier to understand formulas since (a prefix of) a word can occur more than once in a sample. As a consequence, however, the formula $\varphi_n^\Un$ has to be changed in comparison to Section~\ref{sec:learner}.

\subsection*{Recapping the main ideas and encoding of states and transitions}
The key idea of the formula $\varphi_n^\mathcal S$ is to encode a \dfa with $n$ states by means of Boolean variables and to pose constraints on those variables in order to obtain a \dfa that is consistent with the given sample. Our encoding relies on a simple observation: if we fix the alphabet, the set of states and the initial state, then any \dfa with $n$ states is uniquely determined (up to isomorphism) by its transitions and final states. Hence, we can without loss of generality fix the state set of the prospective \dfa to be $Q = \{ 0,  \ldots, n-1 \}$ and the initial state to be $q_0 = 0$; the alphabet $\Sigma$ is determined by the given game.

To encode the transitions and the final states, we introduce Boolean variables $d_{p, a, q}$ and $f_q$ where $p, q \in Q$ and $a \in \Sigma$, which have the following meaning: assigning \true to $d_{p, a, q}$ means that the transition $\delta(p, a) = q$ exists in the prospective \dfa, and assigning \true to $f_q$ means that $q$ is a final state.

To make sure that the variables $d_{p, a, q}$ indeed encode a deterministic transition function, we impose the following constraints.
\begin{align}
	\bigwedge_{p \in Q}~ \bigwedge_{a \in \Sigma}~ \bigwedge_{q, q' \in Q, q \neq q'}~ \lnot d_{p, a, q} \lor \lnot d_{p, a, q'} \label{for:dfa_1} \\
	\bigwedge_{p \in Q}~ \bigwedge_{a \in \Sigma}~ \bigvee_{q \in Q}~ d_{p, a, q} \label{for:dfa_2}
\end{align}
Formula~\eqref{for:dfa_1} and \eqref{for:dfa_2} are the same as Formula~\eqref{main:for:dfa_1} and \eqref{main:for:dfa_2} of Section~\ref{sec:learner}, respectively: Formula~\eqref{for:dfa_1} enforces that $d_{p, a, q}$ encode a deterministic function, while Formula~\eqref{for:dfa_2} asserts that the function is total.

Let $\varphi_n^\text{DFA}(\overline d, \overline f)$ be the conjunction of Formulas~\eqref{for:dfa_1} and \eqref{for:dfa_2} where $\overline d$ denotes the list of variables $d_{p, a, q}$ and $\overline f$ denotes the list of variables $f_q$ for $p, q \in Q$ and $a \in \Sigma$. Given a model $\mathfrak M$ of $\varphi_n^\text{DFA}$, deriving the encoded \dfa is straightforward, as shown next.

\begin{definition}\label{def:conjecture_from_SAT}
Let $\mathfrak M \models \varphi_n^\text{DFA}(\overline d, \overline f)$. We define the \dfa $\mathcal A_\mathfrak M = (Q, \Sigma, q_0, \delta, F)$ by
\begin{itemize}
    \item $\delta(p, a) = q$ for the unique $q \in Q$ with $\mathfrak M(d_{p, a, q}) = \true$; and 
	\item $F = \{ q \in Q \mid \mathfrak M(f_q) = \true \}$.
\end{itemize}
(Recall that we fixed $Q = \{0, \ldots, n-1 \}$ and $q_0 = 0$.)
\end{definition}

To produce a \dfa that is consistent with a sample, we add further constraints:
\begin{itemize}
	\item a formula $\varphi_n^\Po$ asserting $\Po \subseteq L(\mathcal A_\mathfrak M)$;
	\item a formula $\varphi_n^\Ne$ asserting $\Ne \cap L(\mathcal A_\mathfrak M) = \emptyset$;
	\item a formula $\varphi_n^\Ex$ asserting for each $(u, A) \in \Ex$ that $u \in L(\mathcal A_\mathfrak M)$ implies $L(\mathcal A_\mathfrak M) \cap L(\mathcal A) \neq \emptyset$; and
    \item a formula $\varphi_n^\Un$ asserting for each $(u, A) \in \Ex$ that $u \in L(\mathcal A_\mathfrak M)$ implies $L(\mathcal A_\mathfrak M) \subseteq L(\mathcal A)$.
\end{itemize}
Moreover, we add an auxiliary formula $\varphi_n^W$, which we discuss shortly. Then, 
\[ \varphi_n^\mathcal S \coloneqq \varphi_n^\text{DFA} \land \varphi_n^\words \land \varphi_n^\Po \land \varphi_n^\Ne \land \varphi_n^\Ex \land \varphi_n^\Un \]
is the desired formula.

The pivotal idea of these formulas is to impose constraints on the variables $d_{p, a, q}$ and $f_q$, which, in turn, determine the \dfa $\mathcal A_\mathfrak M$. Having this in mind, it is easier to describe the effects of these constraints by referring to $\mathfrak M$ rather then to the variables themselves. However, we thereby implicitly assume that the formula is satisfiable and that the valuation $\mathfrak M$ is a model.

\mysection{The formula \texorpdfstring{$\varphi_n^\words$}{phi\textunderscore{}n\textasciicircum{}W}}
To ensure that the prospective automaton $\mathcal A_\mathfrak M$ is consistent with the given sample, we need a mechanism to determine whether $\mathcal A_\mathfrak M$ accepts or rejects the words occurring in the sample. The idea is to track the run of $\mathcal A_\mathfrak M$ on all prefixes of the set 
\[ \words = \Po \cup \Ne \cup \ante(\Ex) \cup \ante (\Un), \]
which contains all positive and negative words as well as all words that occur as antecedent of an existential or universal implication. The idea is to introduce auxiliary Boolean variables $x_{u, q}$ where $u \in \pref(\words)$ and $q \in Q$; the intended meaning of these variables is that if the prospective \dfa $\mathcal A_\mathfrak M$ reaches state $q$ on reading the word $u$, then $x_{u, q}$ is set to \true. The following constraints enforce this.
\begin{align}
	x_{\varepsilon, q_0} \label{for:run_1} \\
    \bigwedge_{u \in \pref(\words)}~ \bigwedge_{q \neq q' \in Q}~ \neg x_{u, q} \vee \neg x_{u,q'} \label{for:run_2} \\
    \bigwedge_{ua \in \pref(\words)}~ \bigwedge_{p,q \in Q}~ (x_{u, p} \wedge d_{p, a, q}) \rightarrow x_{ua, q} \label{for:run_3}
\end{align}
Since any run starts in the initial state $q_0$, Formula~\eqref{for:run_1} asserts that $x_{\varepsilon, q_0}$ is set to \true. Formula~\eqref{for:run_2} enforces that for every $u \in \pref(W)$ there exists at most one $q \in Q$ such that $x_{u, q}$ is set to \true (in fact, the conjuction of Formulas~\eqref{for:dfa_2}--\eqref{for:run_3} implies that there exists a unique such state). Finally, Formula~\eqref{for:run_3} prescribes how the run of $\mathcal A_\mathfrak M$ on a word $u \in \pref(\words)$ proceeds: if $\mathcal A_\mathfrak M$ reaches state $p$ on reading $u$ (i.e., $x_{u, p}$ is set to $\true$) and there exists a transition from $p$ to state $q$ on reading the symbol $a \in \Sigma$ (i.e., $d_{p, a, q}$ is set to \true), then $\mathcal A_\mathfrak M$ reaches state $q$ on reading $ua$ and $x_{ua}$ is set to \true.

Let $\varphi_n^\words(\overline d, \overline f, \overline x)$ be the conjunction of Formulas~\eqref{for:run_1}, \eqref{for:run_2}, and \eqref{for:run_3} where $\overline d$ and $\overline f$ are as above and $\overline x$ is the list of variables $x_{u, q}$ for $u \in \pref(W)$ and $q \in Q$. Then a stright-forward induction proves the following lemma (see, e.g., Neider and Jansen~\cite{DBLP:conf/nfm/NeiderJ13}).

\begin{lemma}\label{lem:run}
Let $n \geq 1$, $\mathfrak M$ a model of
\[ \varphi_n^\text{DFA}(\overline d, \overline f) \wedge \varphi_n^\words(\overline d, \overline f, \overline x), \]
and $\mathcal A_\mathfrak M$ the DFA defined according to Definition~\ref{def:conjecture_from_SAT}. Then, $\mathcal A_\mathfrak M \colon q_0 \myrightarrow{u} q$ implies $\mathfrak M(x_{u, q}) = \true$ for all $u \in \pref(\words)$.
\end{lemma}

\mysection{The formulas \texorpdfstring{$\varphi_n^\Po$}{phi\textunderscore{}n\textasciicircum{}Pos} and \texorpdfstring{$\varphi_n^\Ne$}{phi\textunderscore{}n\textasciicircum{}Ne}}
Having introduced the formula $\varphi_n^\words$, it is straightforward to enforce a correct behavior of $\mathcal A_\mathfrak M$ on \Po and \Ne. To assert that $\mathcal A_\mathfrak M$ accepts all words in $\Po$, we impose the constraint
\begin{align}
	\bigwedge_{u \in \Po}~ \bigwedge_{q \in Q}~ x_{u, q} \rightarrow f_q, \label{for:pos_1}
\end{align}
which ensures that state $q$ is a final state if $\mathcal A_\mathfrak M$ reaches $q$ on reading a word $u \in \Po$.
Similarly, the constraint
\begin{align}
	\bigwedge_{u \in \Ne}~ \bigwedge_{q \in Q}~  x_{u, q} \rightarrow \neg f_q \label{for:neg_1}
\end{align}
makes sure that state $q$ is not a final state if $\mathcal A_\mathfrak M$ reaches $q$ on reading a word $u \in \Ne$, hence, asserting that all words of $\Ne$ are rejected.

Let $\varphi_n^\Po(\overline d, \overline f, \overline x)$ denote Formula~\eqref{for:pos_1} and $\varphi_n^\Ne(\overline d, \overline f, \overline x)$ denote Formula~\eqref{for:neg_1} where $\overline d$, $\overline f$, and $\overline x$ are as above. Then, we obtain the following results.

\begin{lemma}\label{lem:pos_correct}
Let $\mathcal S = (\Po, \Ne, \Ex, \Un)$ be  a sample, $n \geq 1$, and
\[
	\psi_n^\Po(\overline d, \overline f, \overline x) \coloneqq \varphi_n^\text{DFA}(\overline d, \overline f) \land \varphi_n^\words(\overline d, \overline f, \overline x) \land \varphi_n^\Po(\overline d, \overline f, \overline x).
\]
Then, the following statements hold:
\begin{enumerate}
	\item \label{stmt:pos_correct_a} If $\mathfrak M \models \psi_n^\Po$, then $\mathcal A_\mathfrak M$ is a \dfa with $n$ states that satisfies $\Po \subseteq L(\mathcal A_\mathfrak M)$.
	\item \label{stmt:pos_correct_b} If a \dfa $\mathcal B$ with $n$ states exists that satisfies $\Po \subseteq L(\mathcal B)$, then $\psi_n^\Po$ is satisfiable.
\end{enumerate}
\end{lemma}

\begin{lemma}\label{lem:neg_correct}
Let $\mathcal S = (\Po, \Ne, \Ex, \Un)$ be  a sample, $n \geq 1$, and
\[
	\psi_n^\Ne(\overline d, \overline f, \overline x) \coloneqq \varphi_n^\text{DFA}(\overline d, \overline f) \land \varphi_n^\words(\overline d, \overline f, \overline x) \land \varphi_n^\Ne(\overline d, \overline f, \overline x).
\]
Then, the following statements hold:
\begin{enumerate}
	\item \label{stmt:neg_correct_a} If $\mathfrak M \models \psi_n^\Ne$, then $\mathcal A_\mathfrak M$ is a \dfa with $n$ states that satisfies $\Ne \cap L(\mathcal A_\mathfrak M) = \emptyset$.
	\item \label{stmt:neg_correct_b} If a \dfa $\mathcal B$ with $n$ states exists that satisfies $\Ne \cap L(\mathcal B) = \emptyset$, then $\psi_n^\Ne$ is satisfiable.
\end{enumerate}
\end{lemma}

Let us now prove Lemma~\ref{lem:pos_correct}. The proof of Lemma~\ref{lem:neg_correct} is analogous.

\begin{IEEEproof}[Proof of Lemma~\ref{lem:pos_correct}]
To prove the Statement~\ref{stmt:pos_correct_a}, assume $\mathfrak M \models \psi_n^\Po$ and let $\mathcal A_\mathfrak M$ be the \dfa constructed according to Definition~\ref{def:conjecture_from_SAT}. Furthermore, pick an arbitrary $u \in \Po$. Then, Lemma~\ref{lem:run} implies that if $\mathcal A_\mathfrak M$ reaches state $q$ on reading $u$, then $\mathfrak M(x_{u, q}) = \true$. Additionally, Formula~\eqref{for:pos_1} asserts that $q$ is a final state and, therefore, $\mathcal A_\mathfrak M$ accepts $u$ by Definition~\ref{def:conjecture_from_SAT}. Since this is true for all $u \in \Po$, we obtain $\Po \subseteq L(\mathcal A_\mathfrak M)$.

To  prove the second statement, let $\mathcal B = (Q_\mathcal B, \Sigma, q_0^\mathcal B, \delta_\mathcal B, F_\mathcal B)$ be a \dfa with $n$ states that satisfies $\Po \subseteq L(\mathcal B)$. The key idea is to translate $\mathcal B$ into a valuation $\mathfrak V$ that satisfies $\psi_n^\Po$. To simplify this translation a bit, we assume without loss of generality that the sets of states of $\mathcal B$ and $\mathcal A_\mathfrak M$ coincide (i.e., $Q_\mathcal B = Q$); one can easily achieve this by renaming states. The definition of $\mathfrak V$ is a follows:
\begin{itemize}
	\item For each $p ,q \in Q_\mathcal B$ and $a \in \Sigma$, we set $d_{p, a, q}$ to \true if and only if $\delta_\mathcal B(p, a) = q$.
	\item For each $q \in Q_\mathcal B$, we set $f_q$ to \true if and only if $q \in F_\mathcal B$.
	\item For each $u \in W$, we set $x_{u, q}$ to \true if and only if $\mathcal B \colon q_0^\mathcal B \myrightarrow{u} q$.
\end{itemize}
It is not hard to verify that $\mathfrak V$ indeed satisfies $\psi_n^\Po$ since $\mathfrak V(x_{u, q})$ is defined according to the runs of $\mathcal B$ on the inputs $u \in \words$.
\end{IEEEproof}

\mysection{The formula \texorpdfstring{$\varphi_n^\Un$}{phi\textunderscore{}n\textasciicircum{}Uni}}
The formula $\varphi_n^\Un$ needs to enforce that $L(\mathcal A_\mathfrak M)$ respects all universal implications in $\Un$. (Recall that the learner stores universal and existential implication as a pair $(u, \mathcal A)$ where $u \in \Sigma^\ast$ is a word and $\mathcal A $ is an \nfa over $\Sigma$.)
To achieve this, we construct for each universal implication $\iota = (u, \mathcal A) \in \Un$ a formula $\varphi_n^\iota$ that asserts $L(\mathcal A)\subseteq L(\mathcal A_\mathfrak M)$ if $u \in L(\mathcal A_\mathfrak M)$. The formulas $\varphi_n^\Un$ is then the (finite) conjunction $\bigwedge_{\iota \in \Un} \varphi_n^\iota$.

Given a universal implication $\iota \in \Un$, say $\iota = (u, \mathcal A)$ with $\mathcal A = (Q_\mathcal A, \Sigma, q_0^\mathcal A, \Delta_\mathcal A, F_\mathcal A)$, the key idea of the formula $\varphi_n^\iota$ is to track the runs of $\mathcal A_\mathfrak M$ and $\mathcal A$ in parallel.
To this end, we introduce new auxiliary variables $y_{q, q'}^\iota$ where $q \in Q$ and $q' \in Q_\mathcal A$, which have the following meaning: the variable $y_{q, q'}^\iota$ is set to \true if there exists a word $v \in \Sigma^\ast$ such that $\mathcal A_\mathfrak M \colon q_0 \myrightarrow{v} q$ and $\mathcal A \colon q_0^\mathcal A \myrightarrow{v} q'$. The following constraints assert this.
\begin{align}
	y_{q_0, q_0^\mathcal A}^\iota \label{for:universal_1} \\
	\bigwedge_{p, q \in Q}~ \bigwedge_{(p', a, q') \in \Delta_\mathcal A}~ (y_{p, p'}^\iota \land d_{p, a, q}) \rightarrow y_{q, q'}^\iota \label{for:universal_2}
\end{align}
Formula~\eqref{for:universal_1} enforces $y_{q_0, q_0^\mathcal A}^\iota$ to be set to \true because $\mathcal A_\mathfrak M \colon q_0 \myrightarrow{\varepsilon} q_0$ and $\mathcal A \colon q_0^\mathcal A \myrightarrow{\varepsilon} q_0^\mathcal A$. Formula~\eqref{for:universal_2} is similar to Formula~\eqref{for:run_3} and describes how the runs of $\mathcal A_\mathfrak M$ and $\mathcal A$ proceed: if there exists a word $v$ such that $\mathcal A_\mathfrak M \colon q_0 \myrightarrow{v} p$ and $\mathcal A \colon q_0^\mathcal A \myrightarrow{v} p'$ (i.e., $y_{p, p'}^\iota$ is set to \true) and there are transitions $(p', a, q') \in \Delta_\mathcal A$ and $\delta(p, a) = q$ in $\mathcal A_\mathfrak M$, then $\mathcal A_\mathfrak M \colon q_0 \myrightarrow{va} q$ and $\mathcal A \colon q_0^\mathcal A \myrightarrow{va} q'$, which requires that $y_{q, q'}^\iota$ has to be set to \true as well.

Note that the variables $y_{q, q'}^\iota$ do not track runs exactly: it is possible that a variable $y_{q, q'}$ is set to \true even without the existence of a word $v \in \Sigma^\ast$ that induces the runs $\mathcal A_\mathfrak M \colon q_0 \myrightarrow{v} q$ and $\mathcal A \colon q_0^\mathcal A \myrightarrow{v} q'$. This inaccuracy, however, is sufficient to obtain the desired result.

In order to express that $\mathcal A_\mathfrak M$ indeed respects the universal implication $\iota$, we add the implication
\begin{align}
	\left( \bigvee_{q \in Q}~ x_{u, q} \land f_q \right) \rightarrow \left( \bigwedge_{q \in Q}~ \bigwedge_{q' \in F_\mathcal A}~ y_{q, q'}^\iota \rightarrow f_q \right). \label{for:universal_3}
\end{align}
This formula ensures that whenever $\mathcal A_\mathfrak M$ accepts $u$ (i.e., the antecedent is \true), then all words that lead to an accepting state in $\mathcal A$ also lead to an accepting state in $\mathcal A_\mathfrak M$ (i.e., the consequent is \true).

Let $\varphi_n^\iota(\overline d, \overline f, \overline x, \overline{y^\iota})$ be the conjunction of Formulas~\eqref{for:universal_1}, \eqref{for:universal_2}, and \eqref{for:universal_3} where $\overline d$, $\overline f$, as well as $\overline x$ are as above and $\overline {y^\iota}$ is the list of all $y_{q, q'}^\iota$ for $q \in Q$ and $q' \in Q_\mathcal A$. Additionally, let $\varphi_n^\Ex$ be the conjunction
\[ \varphi_n^\Un (\overline d, \overline f, \overline x, \overline y) \coloneqq  \bigwedge_{\iota \in \Un}~ \varphi_n^\iota(\overline d, \overline f, \overline x, \overline{y^\iota}), \]
where $\overline y$ denotes the list of all variables occurring in $\overline{y^\iota}$ for each $\iota \in \Un$. Then, the following holds.

\begin{lemma}\label{lem:universal_correct}
Let $\mathcal S = (\Po, \Ne, \Ex, \Un)$ be a sample, $n \geq 1$, and
\begin{multline*}
	\psi_n^\Un(\overline d, \overline f, \overline x, \overline y) \coloneqq  \varphi^\text{DFA}_n(\overline d, \overline f) \\
	{} \land \varphi_n^\words(\overline d, \overline f, \overline x) \land \varphi_n^\Un(\overline d, \overline f, \overline x, \overline y).
\end{multline*}
Then, the following statements hold:
\begin{enumerate}
	\item\label{stmt:universal_correct_a} If $\mathcal M \models \psi_n^\Un$, then $\mathcal A_\mathfrak M$ is a \dfa with $n$ states that satisfies for all $(u, \mathcal A) \in \Un$ that $u \in L(\mathcal A_\mathfrak M)$ implies $L(\mathcal A) \subseteq L(\mathcal A_\mathfrak M)$.
	\item\label{stmt:universal_correct_b} If a \dfa with $n$ states exists that satisfies for all $(u, \mathcal A) \in \Un$ that $u \in L(\mathcal A_\mathfrak M)$ implies $L(\mathcal A)  \subseteq L(\mathcal A_\mathfrak M)$, then $\psi_n^\Un$ is satisfiable.
\end{enumerate}
\end{lemma}

\begin{IEEEproof}
We split the proof in two parts: we first show Statement~\ref{stmt:universal_correct_a} and subsequently Statement~\ref{stmt:universal_correct_b}. \par\vspace{\baselineskip}

\noindent To prove Statement~\ref{stmt:universal_correct_a}, we show that for an universal implication $\iota = (u, \mathcal A) \in \Un$, a model $\mathfrak M$ of the formula
\[ \psi^\iota_n(\overline d, \overline f, \overline x, \overline{y^\iota}) \coloneqq \varphi^\text{DFA}_n(\overline d, \overline f) \wedge \varphi_n^\words(\overline d, \overline f, \overline x) \wedge \varphi_n^\iota(\overline d, \overline f, \overline x, \overline{y^\iota}) \]
results in an automaton $\mathcal A_\mathfrak M$ that respects $\iota$ (i.e., $u \in L(\mathcal A_\mathfrak M)$ implies $L(\mathcal A) \subseteq L(\mathcal A_\mathfrak M)$). The claim of Statement~\ref{stmt:universal_correct_a} then follows immediately because $\varphi_n^\Un$ is the conjunction of the individual formulas $\varphi^\iota_n$. In the following, fix an universal implication $\iota = (u, \mathcal A) \in \Un$, assume $\mathfrak M \models \psi_n^\iota$, and let $\mathcal A_\mathfrak M$ be the \dfa constructed according to Definition~\ref{def:conjecture_from_SAT}.

Given an universal implication $\iota = (u, \mathcal A)$, say with $\mathcal A = (Q_\mathcal A, \Sigma, q_0, \Delta_\mathcal A, F_\mathcal A)$, we first show by induction over the length of inputs $v \in \Sigma^\ast$ that the variables $y_{q, q'}^\iota$ have indeed the desired meaning (i.e., $\mathcal A_\mathfrak M \colon q_0 \myrightarrow{v} q$ and $\mathcal A \colon q_0^\mathcal A \myrightarrow{v} q'$ imply $\mathfrak M(y_{q, q'}^\iota) = \true$).

\begin{description}[font=\normalfont\itshape]
	\item[Base case ($v = \varepsilon$)] Both $\mathcal A_\mathfrak M \colon q_0 \myrightarrow{\varepsilon} q_0$ and $\mathcal A \colon q_0^\mathcal A \myrightarrow{\varepsilon} q_0^\mathcal A$ hold by definition of runs. Moreover, Formula~\eqref{for:universal_1} enforces $\mathfrak M(y_{q_0, q_0^\mathcal A}^\iota) = \true$. Thus, the claim holds. 
	\item[Induction step ($v = v'a$)] Assume $\mathcal A_\mathfrak M \colon q_0 \myrightarrow{v'} p \myrightarrow{a} q$ and $\mathcal A \colon q_0^\mathcal A \myrightarrow{v'} p' \myrightarrow{a} q'$. Thus, there exists transitions $(p', a, q') \in \Delta_\mathcal A$ and $\delta(p, a) = q$; the latter means $\mathfrak M(d_{p, a, q}) = \true$ by Definition~\ref{def:conjecture_from_SAT}. Moreover, applying the induction hypothesis yields $\mathfrak M(y_{p, p'}^\iota) = \true$. In this situation, Formula~\eqref{for:universal_2} enforces $\mathfrak M(y_{q, q'}^\iota) = \true$, which proves the claim.
\end{description}

Having established the meaning of the variables $y_{q, q'}^\iota$, it is now straightforward to prove that $\mathcal A_\mathfrak M$ satisfies $L(\mathcal A) \subseteq L(\mathcal A_\mathfrak M)$ if $u \in L(\mathcal A_\mathfrak M)$. If $\mathcal A_\mathfrak M$ accepts $u$, say $\mathcal A_\mathfrak M \colon q_0 \myrightarrow{u} q$ with $q \in F$, then we know that $\mathfrak M(x_{u, q}) = \true$ (by Lemma~\ref{lem:run}) and $\mathfrak M(f_q) = \true$ (by Definition~\ref{def:conjecture_from_SAT}). In this situation, the antecedent of Formula~\eqref{for:universal_3} is satisfied. Thus, its consequent is necessarily satisfied as well because $\mathfrak M$ is a satisfying assignment of $\psi_n^\Un$. This, in turn, ensures that whenever $\mathcal A$ accepts a word $v \in \Sigma^\ast$, say $\mathcal A \colon q_0^\iota \myrightarrow{v} q'$ with $q' \in F_\mathcal A$, then the run $\mathcal A_\mathfrak M \colon q_0 \myrightarrow{v} q$ is also accepting: the induction above shows that $\mathfrak M(y_{q, q'}^\iota) = \true$ and, since the consequent of Formula~\eqref{for:universal_3} ensures that $\mathfrak M(y_{q, q'}^\iota) = \true$ implies $\mathfrak M(f_q) = \true$ for all $q \in Q$ and $q' \in F_\mathcal A$, also $\mathfrak M(f_q) = \true$ holds. Hence, $L(\mathcal A) \subseteq L(\mathcal A_\mathfrak M)$ because $v$ was chosen arbitrarily. Since these arguments are true for all $\iota \in \Un$, the \dfa $\mathcal  A_\mathfrak M$ respects all implications in $\Un$. \par\vspace{\baselineskip}

\noindent
To prove Statement~\ref{stmt:universal_correct_b}, suppose that $\mathcal B = (Q_\mathcal B, \Sigma, q_0^\mathcal B, \delta_\mathcal B, F_\mathcal B)$ is a \dfa with $n$ states that respects all universal implications in $\Un$. Similar to the proof of Lemma~\ref{lem:run}, we translate this \dfa into a assignment $\mathfrak V$ that satisfies $\psi_n^\Un$. For the sake of this translation, we assume without loss of generality that the state stets of $\mathcal B$ and $\mathcal A_\mathfrak M$ coincide (i.e., $Q_\mathcal B = Q$).

The translation is as follows:
\begin{itemize}
	\item For each $p, q \in Q_\mathcal B$ and $a \in \Sigma$, we set $\mathfrak V(d_{p, a, q}) = \true$ is and only if $\delta_\mathcal B(p, a) = q$.
	\item For each $q \in Q_\mathcal B$, we set $\mathfrak V(f_q) = \true$ if and only if $q \in F_\mathcal B$.
	\item For each $u \in W$ and $q \in Q_\mathcal B$, we set $\mathfrak V(x_{u, q}) = \true$ if and only if $\mathcal B \colon q_0^\mathcal B \myrightarrow{u} q$.
	\item For each universal implication  $\iota = (u, \mathcal A) \in \Un$ with $\mathcal A = (Q_\mathcal A, \Sigma, q_0^\mathcal A, \Delta_\mathcal A, F_\mathcal A)$, $q \in Q_\mathcal B$, and $q' \in Q_\mathcal A$, we set $\mathfrak V(y_{q, q'}^\iota) = true$ if a $v \in \Sigma^\ast$ exists such that $\mathcal B \colon q_0 \myrightarrow{v} q$ and $\mathcal A \colon q_0^\mathcal A \myrightarrow{v} q'$.
\end{itemize}

It is not hard to verify that $\mathfrak V$ satisfies $\varphi_n^\text{DFA} \land \varphi_n^W$. To show that is also satisfies $\varphi_n^\Un$, fix a universal implication $\iota = (u, \mathcal A)$, say with $\mathcal A = (Q_\mathcal A, \Sigma, q_0^\mathcal A, \Delta_\mathcal A, F_\mathcal A)$.
We first observe that $\mathfrak V$ satisfies Formulas~\eqref{for:universal_1} and \eqref{for:universal_2} since the variables $y_{q, q'}^\iota$ track the runs of both automata on inputs $v \in \Sigma^\ast$.
Second, if $u \notin L(\mathcal B)$, then $\mathfrak V$ does not satisfy the antecedent of Formula~\eqref{for:universal_3} and, hence, satisfies Formula~\eqref{for:universal_3}.
If $u \in L(\mathcal B)$, on the other hand, consider the runs $\mathcal B \colon q_0^\mathcal B \myrightarrow{v} q$ and $\mathcal A \colon q_0^\mathcal A \myrightarrow{v} q'$ on some input $v \in \Sigma^\ast$. Then, $\mathfrak V(y_{q, q'}^\iota) = \true$ by definition of $\mathfrak V$. Moreover, if $\mathcal A$ accepts $v$ (i.e., $q' \in F_\mathcal A$), then $\mathcal B$ accepts $v$ as well (i.e., $q \in F_\mathcal B$) because $\mathcal B$ respects all implications in $\Un$. Hence, $\mathfrak V(f_q) = \true$ by definition of $\mathfrak V$. Thus, the valuation $\mathfrak V$ satisfies the consequent of Formula~\eqref{for:universal_3} (since $v$ was chosen arbitrary), which implies that $\mathfrak V$ satisfies Formula~\eqref{for:universal_3}. Finally, we note that these arguments are true for each $\iota \in \Un$ and, thus, $\mathfrak V$ satisfies $\varphi_n^\Un$.
\end{IEEEproof}

\mysection{The formula \texorpdfstring{$\varphi_n^\Ex$}{phi\textunderscore{}n\textasciicircum{}Ex}}
The formula $\varphi_n^\Ex$ needs to enforce that $L(\mathcal A_\mathfrak M)$ respects all existential implications in $\Ex$. Similar to the previous formula, we construct for each existential implication $\iota = (u, \mathcal A) \in \Ex$ a formula $\phi_n^\iota$ that asserts $L(\mathcal A_\mathfrak M) \cap L(\mathcal A) \neq \emptyset$ if $u \in L(\mathcal A_\mathfrak M)$. The formulas $\varphi_n^\Ex$ is then the (finite) conjunction $\bigwedge_{\iota \in \Ex} \phi_n^\iota$.

The formulas $\phi_n^\iota$ work similar to the formulas $\varphi_n^\iota$ introduced above. Given an existential implication $\iota = (u, \mathcal A)$, say with $\mathcal A = (Q_\mathcal A, \Sigma, q_0^\mathcal A, \Delta_\mathcal A, F_\mathcal A)$, the key idea is again to track the runs of $\mathcal A_\mathfrak M$ and $\mathcal A$ in parallel. In contrast to $\varphi_n^\Un$, however, it is no longer sufficient to build upon the variables $y_{q, q'}$ as they do not track the runs exactly; recall that $y_{q, q'}$ might be set to \true even without the existence of a word that induces runs to the state $q \in \mathcal A_\mathfrak M$ and $q' \in \mathcal A$. This fact prevents us from enforcing the existence of a word in the intersection $L(\mathcal A_\mathfrak M) \cap L(\mathcal A)$ based on the variables $y_{q, q'}$ (should this be necessary due to $\mathcal A_\mathfrak M$ accepting the antecedent of $\iota$).

We approach this problem by tracking the parallel runs of $\mathcal A_\mathfrak M$ and $\mathcal A$ exactly, exploiting the following simple fact about finite automata.

\begin{observation}\label{obs:parallel_runs}
Let $\mathcal B_1 = (Q_{\mathcal B_1}, \Sigma, q_0^{\mathcal B_1}, \Delta_{\mathcal B_1}, F_{\mathcal B_1})$ and $\mathcal B_2 = (Q_{\mathcal B_2}, \Sigma, q_0^{\mathcal B_2}, \Delta_{\mathcal B_2}, F_{\mathcal B_2})$ be two \nfas. Then, a word $w \in \Sigma^\ast$ with $\mathcal B_1 \colon q_0^{\mathcal B_1} \myrightarrow{w} q$ and $\mathcal B_2 \colon q_0^{\mathcal B_2} \myrightarrow{w} q'$ exists if and only if a word $w' \in \Sigma^\ast$ of length at most $|Q_{\mathcal B_1}||Q_{\mathcal B_2}| - 1$ with $\mathcal B_1 \colon q_0^{\mathcal B_1} \myrightarrow{w'} q$ and $\mathcal B_2 \colon q_0^{\mathcal B_2} \myrightarrow{w'} q'$ exists.
\end{observation}

To see why Observation~\ref{obs:parallel_runs} is true, suppose there exists an input $w \in \Sigma^\ast$ of length greater than $k = |Q_{\mathcal B_1}||Q_{\mathcal B_2}| - 1$ with $\mathcal B_1 \colon q_0^{\mathcal B_1} \myrightarrow{w} q$ and $\mathcal B_2 \colon q_0^{\mathcal B_2} \myrightarrow{w} q'$. Then, there has to be a pair of states occurring in these runs that repeats at least once. The (nonempty) part of $w$ in between this repetition can be removed, resulting in a word $w'$ with $\mathcal B_1 \colon q_0^{\mathcal B_1} \myrightarrow{w'} q$ and $\mathcal B_2 \colon q_0^{\mathcal B_2} \myrightarrow{w'} q'$. By repeating this argument successively, one obtains a word of length less of equal to $k$ that leads to state $q$ in $\mathcal B_1$ and state $q'$ in $\mathcal B_2$.

As Observation~\ref{obs:parallel_runs} shows, it is indeed enough to consider words of length at most $k = n|\mathcal A| - 1$ in order to track the parallel runs of $\mathcal A_\mathfrak M$ and $\mathcal A$ exactly. We do so by means of new auxiliary variables $z_{q, q', \ell}^\iota$ where $q \in Q$, $q' \in Q_\mathcal A$, and $\ell \in \{0, \ldots, k \}$, which have the following meaning: the variable $z_{q, q', \ell}^\iota$ is set to \true if and only if there exists a word $v \in \Sigma^\ast$ with $|v| = \ell$ such that $\mathcal A_\mathfrak M \colon q_0 \myrightarrow{v} q$ and $\mathcal A \colon q_0^\mathcal A \myrightarrow{v} q'$. The following formulas constrain the variables $z_{q, q', \ell}$ as described. 

\begin{align}
	z_{q_0, q_0^\mathcal A, 0}^\iota ~\land~ \bigwedge_{(q, q') \in Q \times Q_\mathcal A \setminus \{ (q_0, q_0^\mathcal A) \}}~ \lnot z_{q, q', 0}^\iota \label{for:existential_1} \\
    \bigwedge_{p, q \in Q}~ \bigwedge_{(p', a, q') \in \Delta_\mathcal A}~ \bigwedge_{\ell \in \{0, \ldots, k-1 \}}~ (z_{p, p', \ell}^\iota \land d_{p, a, q}) \rightarrow z_{q, q', \ell+1}^\iota \label{for:existential_2}
\end{align}
\begin{multline}
	\bigwedge_{q \in Q}~ \bigwedge_{q' \in Q_\mathcal A}~ \bigwedge_{\ell \in \{ 1, \ldots, k \}}~ z_{q, q', \ell}^\iota \rightarrow \\
	 \bigvee_{p \in Q}~ \bigvee_{(p', a, q') \in \Delta_\mathcal A}~ d_{p, a, q} \land z_{p, p', \ell-1}^\iota \label{for:existential_3}
\end{multline}
Formula~\eqref{for:existential_1} makes sure that $z_{q_0, q_0^\mathcal A, 0}^\iota$ is set to \true, whereas all other variables $z_{q, q', 0}^\iota$ are set to \false, since $\mathcal A_\mathfrak M \colon q_0 \myrightarrow{\varepsilon} q_0$ and $\mathcal A \colon q_0^\mathcal A \myrightarrow{\varepsilon} q_0^\mathcal A$ are the only runs on the empty word. Formula~\eqref{for:existential_2} is similar to Formula~\eqref{for:run_3} and describes how the runs of both automata proceed: if there exists a word $v \in \Sigma^\ast$ with $|v| < k$ that induces the runs $\mathcal A_\mathfrak M \colon q_0 \myrightarrow{v} q$ and $\mathcal A \colon q_0^\mathcal A \myrightarrow{v} q'$ (i.e., $z_{q, q', |v|}^\iota$ is set to \true) and there exists transitions $(p', a, q') \in \Delta_\mathcal A$ and $\delta(p, a) = q$ (i.e., $d_{p, a, q}$ is set to \true), then the word $va$ induces the runs  $\mathcal A_\mathfrak M \colon q_0 \myrightarrow{va} q$ and $\mathcal A \colon q_0^\mathcal A \myrightarrow{va} q'$, which implies that $z_{q, q', |va|}^\iota$ has to be set to \true as well. In a similar manner, Formula~\eqref{for:existential_3} prevents $z_{q, q', \ell}^\iota$ from being set to \true if there exists no input of length $\ell$ that leads to the states $q$ in $\mathcal A_\mathfrak M$ and state $q'$ in $\mathcal A$; an exemption to this constraint is the pair of initial states.

Finally, adding the implication 
\begin{align}
	\left( \bigvee_{q \in Q}~ x_{u, q} \land f_q \right) \rightarrow \left( \bigvee_{q \in Q}~ \bigvee_{q' \in F_\mathcal A}~ \bigvee_{\ell \in \{ 0, \ldots, k \}}~ z_{q, q', \ell}^\iota \wedge f_q \right) \label{for:existential_4}
\end{align}
enforces that $L(\mathcal A_\mathfrak M)$ indeed respects the implication $\iota = (u, \mathcal A)$: if $\mathcal A_\mathfrak M$ accepts $u$ (signaled by the antecedent being \true), then there also has to exist an input on which both automata reach final states (indicated by the consequent being set to \true), hence, proving $L(\mathcal A_\mathfrak M) \cap L(\mathcal A) \neq \emptyset$. 

Let $\phi_n^\iota(\overline d, \overline f, \overline x, \overline{z^\iota})$ be the conjunction of Formulas~\eqref{for:existential_1}--\eqref{for:existential_4} where $\overline d$, $\overline f$, and $\overline x$ are as above and $\overline{z^\iota}$ is a list of variables $z_{q, q', \ell}^\iota$ for $q \in Q$, $q' \in Q_\mathcal A$, and $\ell \in \{ 0, \ldots, k \}$. Moreover, let $\varphi_n^\Ex$ be the conjunction
\[\varphi_n^\Ex(\overline d, \overline f, \overline x, \overline z) \coloneqq \bigwedge_{\iota \in \Ex}~ \phi_n^\iota(\overline d, \overline f, \overline x, \overline{z^\iota}), \]
where $\overline z$ denotes the list of all variables occurring in $\overline{z^\iota}$. Then, the following holds.

\begin{lemma}\label{lem:existential_correct}
Let $\mathcal S = (\Po, \Ne, \Ex, \Un)$ be a sample, $n \geq 1$, and 
\[ \psi_n^\Ex(\overline d, \overline f, \overline x, \overline z) \coloneqq  \varphi^\text{DFA}_n(\overline d, \overline f) \wedge \varphi_n^\words(\overline d, \overline f, \overline x) \wedge \varphi_n^\Ex(\overline d, \overline f, \overline x, \overline z). \]
Then, the following statements hold:
\begin{enumerate}
	\item\label{stmt:existential_correct_a} If $\mathcal M \models \psi_n^\Ex$, then $\mathcal A_\mathfrak M$ is a \dfa with $n$ states that satisfies for all $(u, \mathcal A) \in \Ex$ that $u \in L(\mathcal A_\mathfrak M)$ implies $L(\mathcal A_\mathfrak M) \cap L(\mathcal A) \neq \emptyset$.
	\item\label{stmt:existential_correct_b} If a \dfa with $n$ states exists that satisfies for all $(u, \mathcal A) \in \Ex$ that $u \in L(\mathcal A_\mathfrak M)$ implies $L(\mathcal A_\mathfrak M) \cap L(\mathcal A) \neq \emptyset$, then $\psi_n^\Ex$ is satisfiable.
\end{enumerate}
\end{lemma}

\begin{IEEEproof}[Proof of Lemma~\ref{lem:existential_correct}]
This proof is similar to the proof of Lemma~\ref{lem:universal_correct}. Again, we split this proof into two part: we first prove Statement~\ref{stmt:existential_correct_a} and subsequently Statement~\ref{stmt:existential_correct_b}. \par\vspace{\baselineskip}

\noindent
To prove Statement~\ref{stmt:existential_correct_a}, we show that for an existential implication $\iota = (u, \mathcal A) \in \Ex$, a model of the formula
\[ \psi^\iota_n(\overline d, \overline f, \overline x, \overline{z^\iota}) \coloneqq \varphi^\text{DFA}_n(\overline d, \overline f) \wedge \varphi_n^\words(\overline d, \overline f, \overline x) \wedge \phi_n^\iota(\overline d, \overline f, \overline x, \overline{z^\iota}) \]
results in an automaton $\mathcal A_\mathfrak M$ that respects $\iota$ (i.e., $u \in L(\mathcal A_\mathfrak M)$ implies $L(\mathcal A_\mathfrak M) \cap L(\mathcal A) \neq \emptyset$). The claim of Statement~\ref{stmt:existential_correct_a} then follows immediately because $\varphi_n^\Ex$ is the conjunction of the individual formulas $\phi^\iota_n$. In the following, fix an existential implication $\iota = (u, \mathcal A) \in \Ex$, assume $\mathfrak M \models \psi_n^\iota$, let $\mathcal A_\mathfrak M$ be the \dfa constructed according to Definition~\ref{def:conjecture_from_SAT} and $k = n|Q_\mathcal A| - 1$.

We first prove that the variable $z_{q, q', \ell}^\iota$, where $\ell \in \{ 0, \ldots, k \}$, is set to \true if and only if there exists a $v \in \Sigma^\ast$ with $|v| \leq \ell$ such that $\mathcal A_\mathfrak M \colon q_0 \myrightarrow{v} q$ and $\mathcal A \colon q_0^\mathcal A \myrightarrow{v} q'$. This proof proceeds by induction over $\ell$.

\begin{description}[font=\normalfont\itshape]
	\item[Base case $(\ell = 0$)] The empty word $\varepsilon$ is the unique word $v \in \Sigma^\ast$ with $|v| = 0$. By definition of runs, $\mathcal A_\mathfrak M \colon q_0 \myrightarrow{\varepsilon} q_0$ and $\mathcal A \colon q_0^\mathcal A \myrightarrow{\varepsilon} q_0^\mathcal A$. Moreover, Formula~\eqref{for:existential_1} makes sure that $z_{q_0, q_0^\mathcal A, 0}^\iota$ is set to \true, whereas $z_{q, q', 0}^\iota$ is set to \false for all other pairs of states. In addition, Formula~\eqref{for:existential_3} does not restrict any variable in the case $\ell = 0$. Hence, the claim holds.
	\item[Induction step $(\ell = \ell' + 1)$]  To prove the direction from left to right, assume $\mathfrak M(y_{q, q', \ell}^\iota) = \true$. Then, Formula~\eqref{for:existential_3} asserts that there exists a state $p \in Q$ and a transition $(p', a, q') \in \Delta$ such that $\mathfrak M(z_{p, p', \ell'}^\iota) = \true$ and $\mathfrak M(d_{p, a, q}) = \true$ (the latter means that $\mathcal A_\mathfrak M$ contains the transition $\delta(p, a) = q$). In addition, applying the induction hypothesis yields that there exists a word $v' \in \Sigma^\ast$ with $|v'| = \ell'$ such that $\mathcal A_\mathfrak M \colon q_0 \myrightarrow{v} p$ and $\mathcal A \colon q_0^\mathcal A \myrightarrow{v} p'$. Thus, $v = v'a$ is a word of length $\ell$ satisfying $\mathcal A_\mathfrak M \colon q_0 \myrightarrow{v} q$ and $\mathcal A \colon q_0^\mathcal A \myrightarrow{v} q'$, which proves the claim.

To prove the reverse direction, let $v = v'a \in \Sigma^\ast$ be a word of length $\ell$ and assume that $\mathcal A_\mathfrak M \colon q_0 \myrightarrow{v} p \myrightarrow{a} q$ and $\mathcal A \colon q_0^\mathcal A \myrightarrow{v} p' \myrightarrow{a} q'$. Thus, we know that $(p', a, q') \in \Delta_\mathcal A$ and $\delta(p, a) = q$ (the latter implying $\mathfrak M(d_{p, a, q}) = \true$). In addition, applying the induction hypothesis yields $\mathfrak M(z_{p, p', \ell'}) = \true$. In this situation, Formula~\eqref{for:existential_2} enforces that $z_{q, q', \ell}$ has to be set to \true, which proves the claim.
\end{description}

Having established the correct meaning of the variables $z_{q, q', \ell}$, proving that $\mathcal A_\mathfrak M$ satisfies $L(\mathcal A_\mathfrak M) \cap L(\mathcal A) \neq \emptyset $ if $u \in L(\mathcal A_\mathfrak M)$ is now straightforward: If $u \in L(\mathcal A_\mathfrak M)$, say $\mathcal A_\mathfrak M \colon q_0 \myrightarrow{u} q$ with $q \in F$, then we know that $x_{u, q}$ is set to \true (by Lemma~\ref{lem:run}) and that $\mathfrak M(f_q) = \true$ (by Definition~\ref{def:conjecture_from_SAT}). In this situation, the antecedent of Formula~\eqref{for:existential_1} is satisfied, which implies that its consequent is satisfied as well (since $\mathfrak M$ is a model of $\psi_n^\iota$). This means that there exist $q \in Q$, $q' \in F_\mathcal A$, and $\ell \in \{ 0, \ldots, k \}$ such that both $\mathfrak M(z_{q, q', \ell}) = \true$ and $\mathfrak M(f_q) = \true$. The former asserts that there exists a word $v \in \Sigma^\ast$ (of length $\ell$) such that $\mathcal A_\mathfrak M \colon q_0 \myrightarrow{v} q$ and $\mathcal A \colon q_0^\mathcal A \myrightarrow{v} q'$ (according to the induction above); on the other hand, the latter means $q \in F$. Hence $v$ is accepted by both automata and, consequently, $u \in L(\mathcal A_\mathfrak M)$ implies $L(\mathcal A_\mathfrak M) \cap L(\mathcal A) \neq \emptyset$. \par\vspace{\baselineskip}

\noindent
To prove Statement~\ref{stmt:existential_correct_b}, let $\mathcal B = (Q_\mathcal B, \Sigma, q_0^\mathcal B, \delta_\mathcal B, F_\mathcal B)$ be a \dfa with $n$ states that satisfies $L(\mathcal B) \cap L(\mathcal A) \neq \emptyset$ if $u \in L(\mathcal B)$ for all $(u, \mathcal A) \in \Ex$. Similar to the previous proofs, we translate $\mathcal B$ into a satisfying valuation $\mathfrak V$ of the variables $\overline d$, $\overline f$, $\overline x$, and $\overline z$. For the sake of this translation, we once more assume without loss of generality that the sets of states of $\mathcal B$ and $\mathcal A_\mathfrak M$ coincide (i.e., $Q_\mathcal B = Q$). The definition of $V$ then is as follows:
\begin{itemize}
	\item For each $p, q \in Q_\mathcal B$ and $a \in \Sigma$, we set $\mathfrak V(d_{p, a, q}) = \true$ if and only if $\delta_\mathcal B(p, a) = q$.
	\item For each $q \in Q_\mathcal B$, we set $\mathfrak V(f_q) = \true$ if and only if $q \in F_\mathcal B$.
	\item For each $u \in W$ and $q \in Q_\mathcal B$, we set $\mathfrak V(x_{u, q}) = \true$ if and only if $\mathcal B \colon q_0^\mathcal B \myrightarrow{u} q$.
	\item For each $\iota = (u, \mathcal A) \in \Ex$, where $\mathcal A = (Q_\mathcal A, \Sigma, q_0^\mathcal A, \Delta_\mathcal A, F_\mathcal A)$, $q \in Q_\mathcal B$, and $q' \in Q_\mathcal A$, we set $\mathfrak V(z_{q, q', \ell}^\iota) = \true$ if and only if there exists a word $v \in \Sigma^\ast$ with length $\ell \leq n|Q_\mathcal A| - 1$ such that $\mathcal B \colon q_0^\mathcal B \myrightarrow{v} q$ and $\mathcal A \colon q_0^\mathcal A \myrightarrow{v} q'$ .
\end{itemize}

It is not hard to verify that $\mathfrak V$ satisfies $\varphi_n^\text{DFA} \land \varphi_n^W$. To see why it also satisfies $\varphi_n^\Ex$, pick a universal implication $(u, A) \in \Ex$, say with $\mathcal A = (Q_\mathcal A, \Sigma, q_0^\mathcal A, \Delta_\mathcal A, F_\mathcal A)$, and let $k = |Q_\mathcal B||Q_\mathcal A|$ (recall that $|Q_\mathcal B| = n = |Q|$).
 First, it is not hard to see that $\mathfrak V$ satisfies Formulas~\eqref{for:existential_1} to \eqref{for:existential_3} since these formulas exactly describe the runs of $\mathcal B$ and $\mathcal A$ on words of length at most $k$.
Second, if $u \notin L(\mathcal B)$, then $\mathfrak V$ does not satisfy the antecedent of Formula~\eqref{for:existential_4} and, hence, satisfies Formula~\eqref{for:existential_4}.
If $u \in L(\mathcal B)$, on the other hand, we know that $L(\mathcal B) \cap L(\mathcal A) \neq \emptyset$.

In other words, there exists a word $v \in L(\mathcal B) \cap L(\mathcal A)$ such that $\mathcal B \colon q_0^\mathcal B \myrightarrow{v} q$ and $\mathcal A \colon q_0^\mathcal A \myrightarrow{v} q'$ where $q \in F_\mathcal B$ and $q' \in F_\mathcal A$. Moreover, Observation~\ref{obs:parallel_runs} allows us to assume without loss of generality that $|v| \leq k$. In this situation, $\mathfrak V(z_{q, q', |v|}^\iota) = \true$ and $\mathfrak V(f_q) = \true$ holds by definition of $\mathfrak V$. Hence, $\mathfrak V$ satisfies the consequent of Formula~\eqref{for:existential_4}, which implies that $\mathfrak V$ satisfies Formula~\eqref{for:existential_4} as well. Finally, since these arguments are true for each $\iota \in \Ex$, the valuation $\mathfrak V$ satisfies $\varphi_n^\Ex$.
\end{IEEEproof}

\section{Correctness of the SAT Learner}
\label{apdx:sat_learner_proof}

The fact that formula $\varphi_n^\mathcal S$ has the desired properties is a straightforward corollary of Lemmas~\ref{lem:pos_correct} to \ref{lem:existential_correct}.

\begin{corollary}\label{cor:phi_correct}
Let $\mathcal S = (\Po, \Ne, \Ex, \Un)$ be a sample, $n \geq 1$, and 
\begin{multline*}
	\varphi_n^\mathcal S(\overline d, \overline f, \overline x, \overline y, \overline z) \coloneqq {} \varphi^\text{DFA}_n(\overline d, \overline f) \land \varphi_n^\words(\overline d, \overline f, \overline x) \land \varphi_n^\Po(\overline d, \overline f, \overline x) \\ 
	{} \land \varphi_n^\Ne(\overline d, \overline f, \overline x) \land \varphi_n^\Un(\overline d, \overline f, \overline x, \overline y) \land \varphi_n^\Ex(\overline d, \overline f, \overline x, \overline z).
\end{multline*}
Then, the following statements hold:
\begin{enumerate}
	\item\label{stmt:phi_correct_a} If $\mathcal M \models \varphi_n^\mathcal S$, then $\mathcal A_\mathfrak M$ is a \dfa with $n$ states that is consistent with $\mathcal S$.
	\item\label{stmt:phi_correct_b} If a \dfa with $n$ states exists that is consistent with $\mathcal S$, then $\varphi_n^\mathcal S$ is satisfiable.
\end{enumerate}
\end{corollary}

Having established that formula $\varphi_n^\mathcal S$ has the desired properties, we can now show that Algorithm~\ref{main:alg:consistent_dfa} computes a smallest \dfa that is consistent with a given sample.

\begin{theorem}\label{thm:consistent_dfa_learner_correct}
Given a contradiction free-sample $\mathcal S$, Algorithm~\ref{main:alg:consistent_dfa} returns a minimal \dfa (in terms of the number of states) that is consistent with $\mathcal S$. In addition, if a minimal consistent \dfa has $k$ states, then Algorithm~\ref{main:alg:consistent_dfa} terminates after $k$ iterations.
\end{theorem}

\begin{IEEEproof}[Proof of Theorem~\ref{thm:consistent_dfa_learner_correct}]
Theorem~\ref{thm:consistent_dfa_learner_correct} follows directly from the properties of the formula $\varphi_n^S$ (see Corollary~\ref{cor:phi_correct}): Given a sample $\mathcal S$, suppose that a \dfa with $k$ states that is consistent with $\mathcal S$ exists. Then, the formula $\varphi_n^\mathcal S$ is satisfiable for all $n \geq k$. Moreover, if $\mathfrak M \models \varphi_n^\mathcal S$, then $\mathcal A_\mathfrak M$ is a \dfa with $n$ states that is consistent with $\mathcal S$. Since Algorithm~\ref{main:alg:consistent_dfa} increases the parameter $n$ by one in every iteration (starting with $n=1$), the algorithm eventually finds the smallest value for which $\varphi_n^\mathcal S$ is satisfiable (after $k$ iterations) and, thus, a consistent \dfa of minimal size. 
\end{IEEEproof}


We are now ready to prove the correctness of the SAT learner.

\begin{theorem}\label{thm:sat_learner_correct}
Given a teacher for a rational safety game, Algorithm~\ref{main:alg:sat_learner}, equipped with Algorithm~\ref{main:alg:consistent_dfa} to construct conjectures, terminates and returns a (minimal) \dfa accepting a winning set if one exists.
\end{theorem}

\begin{IEEEproof}[Proof of Theorem~\ref{thm:sat_learner_correct}]
Due to the way the teacher answers queries, is is clear that the \dfa returned by the SAT learner accepts a winning set. Thus, it is left to show that the SAT learner terminates (given that a winning set exists) and that its result is of minimal size. To this end, we first make three observations:
\begin{enumerate}
	\item \label{obs:sat_learner_correct:1} The SAT learner never conjectures the same \dfa twice. This is due to the fact that the SAT learner only conjectures \dfas that are consistent with the sample of the iteration in which is was constructed. Moreover, a simple proof by contradiction shows that the conjecture of the current iteration is also consistent with the samples of all previous iterations since a new sample results from adding a counterexample (i.e., a word or an implication) to the sample the previous iteration. Hence, the conjectures $\mathcal A_i$ of iteration $i$ and $\mathcal A_j$ of iteration $j < i$ differ at least on the counterexample added in iteration $j$.  
	\item \label{obs:sat_learner_correct:2} The SAT learner conjectures \dfas that grow monotonically in size. To see why, suppose that conjecture $\mathcal A_{i+1}$ of iteration $i+1$ has less states than the conjecture $\mathcal A_i$ of iteration $i$.
	As argued above, $\mathcal A_{i+1}$ is also consistent with the sample $\mathcal S_i$, but has fewer states than $\mathcal A_i$. This, however, contradicts the fact that Algorithm~\ref{main:alg:consistent_dfa} always constructs consistent \dfas of minimal size (see Theorem~\ref{thm:consistent_dfa_learner_correct}).
	\item \label{obs:sat_learner_correct:3} Any \dfa accepting a winning set is consistent with any sample produces during the learning. In other words, adding counterexamples does not rule out solutions.
\end{enumerate}

Theorem~\ref{thm:sat_learner_correct} can now be proven as follows. Suppose that a winning set exists and let $\mathcal A$ be a smallest \dfa, say with $k$ states, that accepts a winning set. Since no smaller \dfa accepting a winning set exists and due to Observations~\ref{obs:sat_learner_correct:1} and \ref{obs:sat_learner_correct:2}, we know that the SAT learner eventually conjectures a \dfa with at least $k$ states. Towards a contradiction, assume that the SAT learner does not conjecture a \dfa with $k$ accepting a winning set. This means that the learner eventually conjectures a \dfa with more than $k$ states. Then, however, Observation~\ref{obs:sat_learner_correct:3} in connection with the fact that the SAT learner always produces smallest consistent \dfas implies that there exists no \dfa with $k$ states accepting a winning set. This is a contradiction. Hence, the SAT learner eventually conjectures a minimal \dfa accepting a winning set, which passes the teacher's query, and terminates.
\end{IEEEproof}

\section{RPNI learner}
\label{apdx:rpni_learner}
The RPNI learner works in a restricted setting in which every vertex of the arena has a finite (but not necessarily bounded) number of outgoing edges (i.e., $E(\{v\})$ is finite for all $v \in V$). This implies that implication counterexamples are of the form $(u, \mathcal A)$ with $L(\mathcal A)$ being finite. 


The RPNI learner works identical to the SAT learner, but uses a different method to construct a consistent \dfa from a sample. While the SAT learner uses a constraint solver for this task (see Algorithm~\ref{main:alg:consistent_dfa}), the RPNI learner employs a modified version of the popular \emph{RPNI algorithm}~\cite{OncinaGarcia92}, which is a polynomial time heuristic for learning \dfas from positive and negative words (we adapted the RPNI algorithm such that it now learns \dfas not only from positive and negative words but also from existential and universal implications). In contrast to Algorithm~\ref{main:alg:consistent_dfa}, however, the modified RPNI algorithm does not, in general, produce minimal consistent \dfas but is much faster. Hence, we encourage the reader to think of the RPNI learner as a heuristic, which uses a faster means to construct conjectures but can no longer guarantee to terminate given that a winning set exists.

As a preparatory step, we first present the original RPNI algorithm. Then, we show how to modify the RPNI algorithm such that it can handle existantial and universal in addition to positive and negative words. Finally, we present the RPNI learner and 

 \subsection{The RPNI Algorithm} 
The RPNI algorithm is a so-called \emph{passive learning algorithm} for regular languages. It takes two disjoint, finite sets $\Po \subset \Sigma^\ast$ and $\Ne \subset \Sigma^\ast$ as input and constructs a \dfa $\mathcal A$ that satisfies $\Po \subseteq L(\mathcal A)$ and $\Ne \cap L(\mathcal A) = \emptyset$. The algorithm runs in time and space polynomial in $|\pref(\Po \cup \Ne)|$ and, hence, the constructed \dfa can, in general, not be minimal (as the problem it solves is NP-complete, see Gold~\cite{DBLP:journals/iandc/Gold78}). It turns out, however, that the RPNI algorithm often produces ``small'' automata in practice.

The RPNI algorithm operates on given sets $\Po$ and $\Ne$ as follows. It first constructs the prefix-tree acceptor of the set $\Po$ (i.e., the tree-like automaton that accepts exactly the set $\Po$). Then, it successively tries to merges states of this automaton (in a fixed order), where a merge is considered to be successful if the resulting \dfa still rejects all words in $\Ne$. If a merge was successful, RPNI proceeds to merge further states of the resulting automaton. If it was not successful, the merged automaton is discarded and RPNI proceeds with the automaton of the last successful merge. The algorithm stops once there are no more merges left.

For our purpose, it is helpful to view the RPNI algorithm as a concrete instance of a generic state-merging algorithm, which is sketched in pseudo code as Algorithm~\ref{alg:gsm}.\kern-0.06em\footnote{The description here closely follows the more general description by Garg et~al.~\cite{DBLP:conf/cav/0001LMN14}.} In this more abstract setting, the learning algorithm takes a finite collection $\kappa$ of data as input and outputs a \dfa that satisfies a given (decidable) property $p$ (which usually refers to $\kappa$); in the case of RPNI, $\kappa$ is the pair $(\Po, \Ne)$ and the property $p$ states that the resulting \dfa has to accept all words in $\Po$ and to reject all words in $\Ne$.
The pivotal idea of Algorithm~\ref{alg:gsm} is to start with a potentially large initial \dfa that satisfies property $p$ and then reduce its size by merging states, thereby discarding merges that result in a \dfa that violates $p$. Since merging states of a \dfa increase its language, we encourage the reader to think of merging as a means of generalization.

\begin{algorithm}
	\BlankLine
	\KwIn{A collection of data $\kappa$}
	\KwOut{\small A \dfa machine $\mathcal A$ that passes \FnTest{$\mathcal A$}}
	\BlankLine
	$\mathcal A_\mathit{init} = (Q, \Sigma, q_0, \delta, f) \gets \FnInit{\text{$\kappa$}}$\;
	\label{alg:gsm:phase1:start}
	$(q_0, \ldots, q_n) \gets \FnOrder{Q}$\;
	\label{alg:gsm:phase1:end}
	\BlankLine
	$\sim_0 \gets \{(q,q) \mid q \in Q \}$\;
	\label{alg:gsm:phase2:start}
	\For{$i = 1, \ldots, n$}{
		\uIf{$q_i \not\sim_{i-1} q_j \text{ for all } j \in \{0, \ldots, i-1\}$}
		{
			$j \gets 0$\;
			\Repeat{\FnTest{$\mathcal{A}_\mathit{init}/_\sim$}}{
				Let $\sim$ be the smallest congruence that contains $\sim_{i-1}$ and the pair $(q_i, q_j)$\;
				$j \gets j+1$\;
			}
			$\sim_i \gets \sim$\;
		}
		\uElse
		{
			$\sim_i \gets \sim_{i-1}$\;
		}
	}
	\label{alg:gsm:phase2:end}
	
	\BlankLine
	\Return $\mathcal A_\mathit{init}/_{\sim_n}$\;
	\caption{Generic state-merging algorithm}
	\label{alg:gsm}
\end{algorithm}

Algorithm~\ref{alg:gsm} uses three functions \FnInit, \FnOrder, and \FnTest, which have the following effects:
\begin{itemize}
	\item The function \FnInit receives a finite collection of data as input and returns a (potentially large) \dfa that satisfies property $p$ (assuming that this is possible).
	\item The function \FnOrder receives a finite set $Q$ as input and returns an ordered sequence of the elements of $Q$.
	\item The function \FnTest receives a \dfa as input and returns a Boolean value indicating whether this \dfa satisfies property $p$.
\end{itemize}
(We shortly introduce implementations of these functions that allows us to compute a \dfa that is consistent with a given finite sample.)

Algorithm~\ref{alg:gsm} runs in two consecutive phases. In the first phase (Lines~\ref{alg:gsm:phase1:start} and \ref{alg:gsm:phase1:end}), it calls the function \FnInit with parameter $\kappa$ to construct an initial \dfa $\mathcal A_\mathit{init}$ that satisfies $p$ (recall that we assume that this is possible). Then, it fixes an order $q_0, \ldots, q_n$ of the states of $\mathcal A_\mathit{init}$ by calling the function \FnOrder with parameter $Q$.

The actual merging takes place in the second phase (Lines~\ref{alg:gsm:phase2:start} to \ref{alg:gsm:phase2:end}), according to the order determined in the first phase. For $i = 1, \ldots, n$ and $j = 0, \ldots, i-1$, the algorithm tries to merge state $q_i$ with state $q_j$ if state $q_i$ has not already been merged with a smaller state; since a merge might introduce nondeterminism, the algorithm merges additional states until determinism is restored. Note that we represent merging of states abstractly as constructing a congruence relation $\sim \subseteq Q \times Q$ (i.e., an equivalence relation that is compatible with the transition function) and the result of the merging as the quotient automaton $\mathcal A_\mathit{init}/_\sim$, which is defined in the usual way. A merge is kept only if the resulting automaton passes \FnTest (otherwise it is discarded). This preserves the invariant that any intermediate \dfa $\mathcal A_\mathit{init}/_{\sim_k}$ satisfies property $p$ (since $\mathcal A_\mathit{init}/_{\sim_0} = \mathcal A_\mathit{init}$ satisfies $p$ by definition of \FnInit). Hence, the final \dfa is guaranteed to satisfy $p$ as well.

\subsection{Adapting the Generic State Merging Algorithm}
\label{sec:adapting_gsm}
In our setting, the collection $\kappa$ corresponds to a sample $\mathcal S =(\Po, \Ne, \Ex, \Un)$, and the property $p$ is consistency with $\mathcal S$. We now describe how to implement the functions \FnInit, \FnOrder, and \FnTest such that the output of Algorithm~\ref{alg:gsm} is a \dfa that is consistent with the input-sample $\mathcal S$.

\paragraph{Creating an initial \dfa}
Given a sample $\mathcal S$, we need to construct a \dfa satisfying $p$ (i.e., a \dfa consistent with $\mathcal S$). To this end, we follow the idea of the RPNI algorithm, namely to construct the prefixtree acceptor of the set $\Po$. The prefix tree acceptor of a finite set $X \subset \Sigma^\ast$ is a partial \dfa \footnote{A \dfa is called \emph{partial} if not all transitions are defined. Runs that cannot be continue due to missing transition are considered to be rejecting.} that accepts exactly the set $X$. It is defined as follows.

\begin{definition}
Given an alphabet $\Sigma$ and finite set $X \subseteq \Sigma^\ast$, the \emph{prefix tree acceptor} is the partial \dfa $\mathcal A_X = (Q, \Sigma, q_0, \delta, F)$ defined by
\begin{itemize} 
	\item $Q = \pref(X)$;
	\item $q_0 = \varepsilon$;
	\item $F  = X$; and
	\item $\delta(u, a) = \begin{cases} ua & \text{if $ua \in \pref(X)$ and}; \\ \text{undefined} & \text{otherwise}. \end{cases}$
\end{itemize}
\end{definition}
A straightforward induction over the length of input-words proves $L(\mathcal A_X) = X$.

However, just starting with the prefix tree acceptor $\mathcal A_\Po$ is not sufficient as $\mathcal A_\Po$ is not necessarily consistent with $\mathcal S$: an implication $(u, \mathcal A)$ might require to accept a word $v \in L(\mathcal A)$ (because $u \in L(\mathcal A_\Po)$) that is not an element of $\Po$ and, hence, is rejected by $\mathcal A_\Po$. In the case of universal implications, the problem is easy to resolve by (temporarily) adding $L(\mathcal A)$ to $\Po$ (recall that $L(\mathcal A)$ is finite). However, the problem becomes more involved in the presence of existential implications as it is no longer apparent which word $v \in L(\mathcal A)$ one should add to $\Po$ in order to obtain a consistent (and preferable small) prefix tree acceptor.

We approach this problem by using a straightforward translation into a satisfiability problem of formulas in propositional Boolean logic (the resulting satisfiability problem is much simpler than those generated by the SAT learner as it does not involve finding a minimal solution). Given a sample $\mathcal S = (\Po, \Ne, \Ex, \Un)$, we introduce a Boolean variable $x_w$ for each 
word $w$ of the set
\begin{multline*}
	V = \Po \cup \Ne \cup \ante(\Ex) \cup \ante(\Un) \\
	{} \cup \left( \bigcup_{(u, \mathcal A) \in \Ex} L(\mathcal A) \right) \cup \left( \bigcup_{(u, \mathcal A) \in \Un} L(\mathcal A) \right),
\end{multline*}
which consists of all words occurring (explicitly and implicitly) in $\mathcal S$. Since the languages of the automata occurring in $\mathcal S$ is finite, $V$ is a finite set and, hence, the number of variables is finite as well.

The desired meaning of the variables is the following: $x_w$ is set to \true if $w$ either belongs to $\Po$ or it is needs to be added to $\Po$ in order to satisfy the implications. The following constraints enforce this meaning.

\begin{align}
	\left( \bigwedge_{w \in \Po} x_w \right) \land \left( \bigwedge_{w \in \Ne} \lnot x_w \right) \label{for:phi:consistent:1} \\
	\bigwedge_{(u, \mathcal A) \in \Ex} \left( x_u \Rightarrow \bigvee_{v \in L(\mathcal A)} x_v \right) \label{for:phi:consistent:2} \\
	\bigwedge_{(u, \mathcal A) \in \Un} \left( x_u \Rightarrow \bigwedge_{v \in L(\mathcal A)} x_v \right) \label{for:phi:consistent:3}
\end{align}

Let $\chi(\overline x)$ be the conjunction of Formulas~\eqref{for:phi:consistent:1}, \eqref{for:phi:consistent:2}, and \eqref{for:phi:consistent:3} where $\overline x$ is the list of all variables $w \in V$. Then, $\chi(\overline x)$ is satisfiable since we assume any sample to be  contradiction-free. Moreover, if $\mathfrak M$ is a model of $\chi(\overline x)$, then the prefix tree acceptor $\mathcal A_{\Po'}$ of the set
\[ \Po' = \{ w \in V \mid \mathfrak M(w) = \true \} \]
is consistent with $\mathcal S$ (i.e., satisfies $p$), which is formalized by the lemma below. This automaton is what the function \FnInit returns.

\begin{lemma}\label{lem:prefix_tree}
Let $\mathcal S =(\Po, \Ne, \Ex, \Un)$ a contradiction-free sample. Then, the following holds:
\begin{enumerate}
	\item The formula $\chi(\overline x)$ is satisfiable.
	\item If $\mathfrak M$ a model of $\chi(\overline x)$ and
	\[ \Po' = \{ w \in V \mid \mathfrak M(w) = \true \}, \]
	then the prefix tree acceptor $\mathcal A_{\Po'}$ is consistent with $\mathcal S$.
\end{enumerate}
\end{lemma}

\begin{IEEEproof}[Proof of Lemma~\ref{lem:prefix_tree}]
Since $\mathcal S$ is contradiction-free, there exists a \dfa, let us denote it by $\mathcal B$, that is consistent with $S$. If we assign \true to the variable $x_w$ if and only if $w \in L(\mathcal B)$, then this assignment satisfies $\chi(\overline x)$. This proves the first claim.

The proof of the second claim relies on the fact that the prefix tree acceptor of a set $X \subseteq \Sigma^\ast$ indeed accepts exactly the set $X$, which can be shown by a simple induction. Given this fact, we first observe that $\mathcal A_{\Po'}$ accepts all words in $\Po$ since  $L(\mathcal A_{\Po'}) = \Po'$ and Formula~\eqref{for:phi:consistent:1} ensures that $\Po \subseteq \Po'$; moreover, a similar argument shows that $\mathcal A_{\Po'}$ rejects all words in $\Ne$. Second, Formula~\eqref{for:phi:consistent:2} asserts for each existential implication $(u, \mathcal A) \in \Ex$ that $u \in \Po'$ implies the existence of a $v \in L(\mathcal A)$ with $v \in \Po' = L(\mathcal A_{\Po'})$. Hence, $\mathcal A_{\Po'}$ respects all existential implications. Moreover, one can establish the fact that $\mathcal A_{\Po'}$ respects all universal implications in an analogous manner by referring to Formula~\eqref{for:phi:consistent:3}.
\end{IEEEproof}

\paragraph{Choosing the Merging Order}
The function \FnInit returns a \dfa whose set of states consists of words over the alphabet $\Sigma$. The order function \FnOrder takes this set and orders it according to the canonical order of words\footnote{Given an alphabet $\Sigma$ and a total order $<_\Sigma \subset \Sigma \times \Sigma$, the \emph{canonical order of words} $\prec \subset \Sigma^\ast \times \Sigma^\ast$ is defined by $a_1 \ldots a_m \prec b_1 \ldots b_n$ if and only if $m < n$ or there exists an $i \in \{1, \ldots, m \}$ such that $a_i <_\Sigma b_i$ and $a_j = b_j$ for all $j \in \{ 1, \ldots, i-1 \}$.}\kern-0.06em. This order is also used by RPNI.

\paragraph{Implementing the Test}
The function \FnTest needs to check whether a given automaton $\mathcal A$ is consistent with the finite sample $\mathcal S$. Since $\mathcal S$ is a finite a collection of words, consistency can be decided easily by computing the runs of $\mathcal A$ on those words and checking whether all four conditions (i.e., acceptance of all words in $\Po$, rejection of all words in $\Ne$, and respecting both types of implications) are fulfilled.

\subsection{Correctness of the RPNI learner}
The correctness of the RPNI learner relies on the  correctness of Algorithm~\ref{alg:gsm}, which is stated in the next lemma.

\begin{lemma}\label{lem:gsm_correct}
Given a contradiction-free sample $\mathcal S$, Algorithm~\ref{alg:gsm} modified as described in Appendix~\ref{sec:adapting_gsm} constructs a \dfa that is consistent with $\mathcal S$. The resulting automaton comprises at most $|V|$ states.
\end{lemma}

\begin{IEEEproof}[Proof of Lemma~\ref{lem:gsm_correct}]
Proving that Algorithm~\ref{alg:gsm} constructs a \dfa that is consistent with the given sample $\mathcal S$ is straightforward: the function \FnInit constructs an initial \dfa that is consistent with $\mathcal S$ (see Lemma~\ref{lem:prefix_tree}), and a merge is only kept if the merged \dfa passes the check \FnTest (i.e., it is still consistent); hence, the final \dfa is guaranteed to be consistent as well. Since the initial \dfa has $|V|$ states and merging of states reduces the number of states, the final \dfa has at most $|V|$ states.
\end{IEEEproof}

The correctness of the RPNI learner immediately follows from the fact that the learning terminates only if the learner proposes a \dfa accepting a winning set. In contrast to the SAT learner, however, the RPNI learner uses an algorithm to derive conjectures that does not necessarily produce consistent \dfas of minimal size. As a consequence, termination of the RPNI learner is not guaranteed even if a \dfa accepting a winning set exists. The following theorem summarizes the main result.

\begin{theorem}\label{thm:rpni_learner_correct}
Given a teacher for a rational safety game over a finitely branching arena, the RPNI learner (i.e., Algorithm~\ref{main:alg:sat_learner} equipped with Algorithm~\ref{alg:gsm} to construct conjectures) on termination returns a \dfa accepting a winning set.
\end{theorem}

\end{document}